\documentclass[prper,reprint,prper]{revtex4-2} 

\usepackage{caption}
\usepackage{amsmath} 
\usepackage{amsfonts} 
\usepackage{graphicx} 
\usepackage{tabularx}
\usepackage[table]{xcolor}
\usepackage{array}
\usepackage{siunitx}
\usepackage{multirow}
\usepackage{tabularx,dcolumn,booktabs}
\usepackage{url}
\usepackage{makecell}

\usepackage{pifont}

\begin{document}

	\title{\emph{Energy} as a source of pre-service teachers' conceptions about radioactivity}

	\author{Axel-Thilo Prokop}
	\email{a.prokop@physik.uni-stuttgart.de}
	\author{Ronny Nawrodt}
	\affiliation{5. Physikalisches Institut, University of Stuttgart, Pfaffenwaldring 57, 70569 Stuttgart, Germany}

	\date{December 2, 2023}
	
	\begin{abstract}
		
		Although researchers have extensively studied student conceptions of radioactivity, the conceptions held by pre-service teachers on this subject are largely absent from the literature. We conducted a qualitative content analysis of problem-centered interviews with pre-service teachers $(N = 13)$ to establish which conceptions are held by pre-service teachers and to examine these conceptions' structure in coordination classes. As has already been observed in students, some pre-service teachers inadequately differentiate between radioactive matter and ionizing radiation and between fission and decay. We also observed that pre-service teachers tend to describe the activation of materials due to ionizing radiation despite having previously denied an activation, thus showing that the conception of activation of materials can reemerge in particular framings. Within the interviews conducted, the concept of energy emerged as a central coordination class regarding radioactivity. This coordination class appeared across contexts and proved fruitful in explaining pre-service teachers' conceptions about radioactivity. We will use the results from this study to develop a teaching-learning laboratory for pre-service teachers in which they can actively study high school students' conceptions while reflecting on their own. In this way, these findings will contribute to improving the structure of nuclear physics courses at the university.

	\end{abstract}
	
	\maketitle 

	\section{INTRODUCTION} 
	
	\emph{Radioactivity} is a physical term that is often familiar to laypeople and students even before they enter any form of formal education on the topic, especially due to its association with nuclear technology. Fundamental to commonly held conceptions is the conflation of radioactive matter with ionizing radiation \cite{Riesch.1975, Eijkelhof.1990}. Further, high school students often fail to distinguish between ionizing and non-ionizing radiation \cite{Riesch.1975, Eijkelhof.1990}. Radioactivity is even associated with its alleged use in non-nuclear technology \cite{Cooper.2003}. Eijkelhof, Henriksen, and Riesch found that people differentiate inadequately between nuclear fission and decay processes.\cite[]{ Eijkelhof.1990, Henriksen.1996, Riesch.1975}. 
	
	While most studies focus on high-school students, the conceptions of laypeople are also a focus of some research and bear similarities to the results of high-school students \cite{Alsop.2001, Henriksen.1996}. In addition to this, the descriptions of half-life are of interest to the research on conceptions about radioactivity \cite{Prather.2005}. The concepts related to radioactivity are explained clearly for both students and non-experts. However, except for the work of Colclough \cite{Colclough.2011}, research on pre-service teachers has been largely overlooked.

	Colclough's study found that pre-service teachers need more clarification on key radiation-related concepts, such as the nature of radioactive decay and the relationship between ionizing radiation and biological effects. Another study from Eijkelhof \cite{Eijkelhof.1990} found that students often have difficulty understanding the concept of risk wh en it comes to ionizing radiation. This is often due to a lack of understanding of the interaction of radiation with matter.

	University students hold various conceptions because of the ubiquity of the term \emph{radioactivity}.  In order to understand and deal with these conceptions and their impacts on teaching radioactivity, it is first necessary to explore them. It is important to note that conceptions are usually not experienced in isolation, but rather in a particular context. To understand how a reactor works, pre-service teachers must piece together various aspects of nuclear physics (\emph{e.g.} the release of energy through fission and the resulting radioactive fission products). This network of discrete pieces of information highlights the need to understand how conceptions are structurally connected. The investigation of an overall structure regarding theories of conceptual change and possible context-dependency in radioactivity is part of the exploratory approach this work takes.

	This paper aims to examine pre-service teachers' conceptions of radioactivity. We will use the conceptions to develop a teaching-learning lab. In the course that prepares pre-service teachers for the teaching-learning lab, the high-school students' conceptions are the basis for their own recognition of these conceptions. This design encourages the teaching-learning lab to function in a way that is rooted in conceptual change.

	\section{Conceptual change}
	
	Conceptual change is widely recognized as a significant part of learning, and several theoretical models have been suggested to clarify it. According to Vosniadou, naïve physics represents a primordial human understanding of physics or natural science \cite{Vosniadou.2013}. Conceptual change is described here as the formation of synthetic models, which can be interpreted as applied theories or as theory-like. This synthesis can be observed in increasingly sophisticated descriptions of Earth's structure by children as they progress through school grades \cite{Vosniadou.2013}. Chi differentiates these applied theories in terms of ontological categories \cite{Chi.2013}. The ontological categories ``mental states'', ``entities'', and ``processes'' form the origin of distinctive categorizations. Additional subdivisions can be classified based on the categories, similar to how a phylogenetic tree is first divided into the kingdoms of life before being organized further into phyla and families. According to Chi, a categorization within the ontological tree leads to stable conceptions \cite{Chi.2013}. For example, the ontological miscategorization of electric current as a substance or entity rather than a process would explain a stable conception. Particular attention is given to the description of emergent processes, as these are present in many alternative conceptions \cite{Chi.2005}. The reassignment of conceptions into their respective exclusive categories is the main task of conceptual change \cite{Chi.2013}. This approach successfully describes students' conceptions of half-life \cite{Hull.2021}.

	DiSessa, unlike Vosniadou and Chi, structures conceptions based on phenomenological primitives (p-prims) and coordination classes, which make up what he describes as a ``conceptual ecology'' \cite{DiSessa.1993,DiSessa.2002,DiSessa.2008}. P-prims represent specific empirical values phenomenologically. For example, one can refer to the idea that ``heavier bodies fall faster'' without questioning it in everyday life. P-prims comprise a toolkit with which students can make constructions \emph{ad-hoc} \cite{DiSessa.1993}. Importantly, the activation of possible elements can be highly dependent on the contexts in which they are applied \cite{DiSessa.2002}. According to DiSessa, structurally related concepts can be elegantly described using coordination classes, which are models that capture the central properties of expert concepts \cite{DiSessa.2005}. Coordination classes are responsible for the ways in which students interpret information in context. The interpretation is guided by gathering and inferring information beyond the context into different situations, forming the causal network. Coordination classes should be able to read or infer information from a broad span of situations and need to be integrated into problem solving. The span, integration, and alignment towards many contexts grant coordination classes their explanatory power \cite{DiSessa.2002}. The importance of prior knowledge arises from the fact that concepts are not necessarily applied where they should be or that their application leads to ``incorrect'' predictions. For example, Levrini and DiSessa \cite{Levrini.2008} state that the concept of force may be applied to a thrown ball but not to a book resting on a table. The question of the breadth of application or predictive ability is called concept projection. In learning, concept projection shows that the application of concepts (such as force) must be learned in different use cases \cite{DiSessa.2005, Levrini.2008}.

	Regarding radioactivity, it is questionable whether a naïve theory can be formed as posed by Vosniadou since there are no comparable processes in everyday experience. Radioactivity is a physical concept \emph{sui generis}. Therefore, a hard distinction of ontological categories in describing the phenomenon of ``radioactivity'' is problematic. Let us consider, for example, the ability of ionizing radiation to interact with matter, which is used as a prompt later on in this study. The distinction of these mechanisms is based on a multitude of properties of the emitted particles and, thus, on many ontological distinctions. For example, alpha and gamma radiation differ in their mass, charge, and interaction mechanisms, and these concepts also interact in statistical processes. A clear misclassification of ontological categories will not simply explain ``misconceptions'' occurring in nuclear physics in general due to its multilevel structure in an ontological sense. As an example, consider nuclear fission, which is a process in the ontological sense. An event triggers this process, making it a direct process. However, at a closer look, the reactivity depends on further quantities like cross-sections. In contrast to Chi, Gupta \emph{et al.} describe that experts act ontologically flexibly \cite{Gupta.2010}. For the concept of energy, diSessa explains how physicists often use it with a substance-like ontological assignment \cite{DiSessa.1993b}. In the context of this work, the aim is to get a glimpse of ``conceptual ecology'' within the topic of radioactivity. To do this, we use ontological classifications to describe the observed ideas. However, the broad choice of topics in the interviews leads to a theoretical classification only being outlined here.

	\section{Conceptions of students regarding radioactivity}
	
	\subsection{Radioactive matter and ionizing radiation}

	Based on qualitative interviews with students, Riesch and Westphal \cite{Riesch.1975} showed that the concept of radiation in the context of radioactivity is associated with mass transport. Many students did not have a clear understanding of radiation as a geometric concept, leading them to interpret the radiation from radioactivity as mass transportation. According to Riesch and Westphal, the terms \textit{radiation} and \textit{radioactivity} are strongly linked \cite{Riesch.1975}. Contrary to today's terminology, Riesch and Westphal spoke of ``radioactive'' radiation, which was standard at the time. The characterization of ``radioactive'' radiation was described using typical physical concepts (waves, gas, rays). A clear separation of the terms ``atom'', ``element'', and ``nucleus'' was not given, and the terms were used synonymously \cite{Prather.2005}. Riesch and Westphal observed that mentioning certain models does not necessarily lead to a consistent mental implementation of them \cite{Riesch.1975}. Mass transport, which relates to radioactivity, was oriented to the spreading of invisible gas, possibly due to the knowledge of the existence of radiogenic radon from the Earth's crust \cite{Boyes.1994, Eijkelhof.1990}.

	The interaction of ionizing or ``radioactive'' radiation with matter directly connects to the perceived conception that radiation is preserved or stored. As per Eijkelhof's research \cite{Eijkelhof.1990}, students tend to assume that food treated with ionizing radiation to disinfect it becomes radioactive itself. In simpler terms, the issue boils down to not being able to distinguish between contamination and irradiation \cite{Eijkelhof.1990}. The central finding that high school students insufficiently differentiate between radioactive matter, ionizing radiation, and the effects of ionizing radiation has been supported by numerous studies \cite[\emph{e.g.}][]{Boyes.1994, Eijkelhof.1990}. In the context of qualitative classroom observations, Eijkelhof further noted that students often fail to differentiate between the concepts of nuclear fission and nuclear disintegration \cite{Eijkelhof.1990}. The lack of distinction between radiation and radioactive material is the best-known conception concerning radioactivity; therefore, the question arises as to whether this conception persists in experienced pre-service teachers.

	\subsection{Half-life}
	
	The temporal description of radioactive processes is closely related to the concept of half-life. Understanding half-life as a characteristic quantity of these processes is part of the study of learners' conceptions of the topic of radioactivity. According to Prather, students often describe the half-life as deterministic \cite{Prather.2005}. Expanding on this notion, the decay of a macroscopic object is associated with a significant loss of mass or volume \cite{Hull.2020, Prather.2005}. Jansky \cite{Jansky.2019} holds that this understanding of half-life is attributed to ignorance, as with other stochastic quantities. This means that while a more exact description is currently beyond our knowledge, it is principally considered to be deterministically explainable. The decay of a nucleus is seen as a continuous process in which the nucleus literally decays \cite{Hull.2021}. The term decay, or \emph{Zerfall} in German (lit. ``disintegration''), suggests that there is a continuous ``decay'' \cite{Hull.2021, Woithe.2017}. An interpretation of this understanding of half-life is carried out by Hull \emph{et al.} \cite{Hull.2021} by considering ontological categories following Chi's reasoning. In doing so, Hull \emph{et al.} argue, that deterministic patterns of interpretation likewise emerge when considering half-life \cite{Hull.2021}. The systematization of half-life is the most consistently implemented conception within the theoretical framework of conceptual change. This highlights the need to study whether pre-service teachers have problems describing decay as an emergent process. Complementary to this, the question arises of how other areas within radioactivity can be integrated into the theoretical framework of conceptual change.

	\subsection{Interaction of ionizing radiation with matter}
	
	As with the description of half-life, the interaction of ionzing radiation with matter is also often subject to deterministic classification. Due to the insufficient differentiation of irradiation and contamination, students describe food irradiation as having unspecified harmful effects \cite{Eijkelhof.1990}. Concerning protection from ionizing radiation, students refer to necessary safety mechanisms, such as maintaining a safe distance \cite{Klaassen.1990}. Students describe the effect on the human body using mechanistic analogies, \emph{e.g.} the disintegration of cells. They primarily consider the harmful impact of ionizing radiation as happening to cells or organs, referring to non-localized defects in genetic material or to general organ failure, respectively \cite{Boyes.1994}. Students' descriptions of ionizing radiation or radioactivity often contain vague references to chemical or technical hazards references. There are clear indications that students interpret radioactivity in a vague sense as a ``chemical'' phenomenon \cite{Eijkelhof.1990}. This loose reference is indicated by the description of toxicity, \emph{e.g.} in the observed description of radioactivity as a gas-like phenomenon \cite{Boyes.1994}. Complementarily, the idea that radioactivity spreads like an infection also appears \cite{Eijkelhof.1990}. 
	
	In the context of the conservation hypothesis, students explain that objects can store radiation \cite{Eijkelhof.1990, Boyes.1994}. The ionization process associated with ionizing radiation and subsequent reactions in matter thus represents a desideratum of conception research. Colclough \emph{et al.} \cite{Colclough.2011} report that pre-service teachers attribute the absorption of ionizing radiation to the density of the absorbing material, which is thought to decrease the count rate within an experiment. The different penetration depths in a material are attributed by students to the differences in energy content of alpha, beta, and gamma radiation \cite{Colclough.2011}. The description of the interaction of radiation with matter is relevant concerning possible measurement methods and the description of possible effects of ionizing radiation on matter. For example, the description of cross-sections is part of the content of the lecture attended by the pre-service teachers described later. Conceptions may persist throughout education, from school to successful completion of the teaching degree.
	
	\subsection{Radioactivity and Technology}
	
	Many students tend to associate radioactivity with technological gadgets, such as radios and cell phones \cite{Boyes.1994,Neumann.2012}. The link between technical devices and radiation is often attributed to human-made pollution causing the release of radioactivity \cite{Boyes.1994}. An important place for students for the release of radioactivity is the nuclear power plant, which is supposed to occur continuously \cite{Boyes.1994}. The general concepts of radiation and radioactivity are deeply connected in the eyes of students \cite{Rego.2006}. Thus, the students' attributions of danger to radiation are related to the statements made regarding radioactivity. Central to these conceptions is the idea that radiation is purely artificial, largely due to the fact that many students tend to associate radioactivity with technological gadgets and human pollution \cite{Neumann.2012}. Consideration of the general concept of radiation in connection with radioactivity also reveals that ``radioactive radiation'' in this sense, is of artificial origin and is increasingly attributed to nuclear power plants, for example. In contrast, the natural occurrence of radioactive processes or ionizing radiation is partly unknown \cite{Rego.2006}. Students often consider radioactivity to be dangerous due to its potential to cause mutations in genetic material. This perception is largely due to the penetrability of radioactivity, which is a notable characteristic of which students are aware. Boyes and Stannisstreet \cite{Boyes.1994} have noted a new classification that emphasizes ionizing radiation's mutagenic or teratogenic properties and provides a clearer understanding of the hazard's cause. Moreover, in a study of Australian students, these ideas persisted despite the intervention. However, teaching within the subject area could help clarify specific uses of radioactive materials such as radiopharmaceuticals \cite{Cooper.2003}. Students always connect radioactivity to its technical application. Numerous application cases (such as radiopharmaceuticals, food irradiation, or nuclear power plants) represent scenarios pre-service teachers can be introduced into their lessons. A greater understanding of to what extent students and pre-service teachers differ in evaluating applications of ionizing radiation or radioactive matter would greatly benefit teaching practices.

	\section{Research questions}
	
	Student conceptions, which are not limited to just students, are also present in other groups like pre-service teachers. This is true in other subfields (\emph{e.g.} mechanics \cite{Halloun.1985}), so we predict it is also the case for radioactivity. The ideas laypeople have about radioactivity are similar in essential to those which (school) students have. Studies of medical students, pre-service teachers, and laypeople have shown that these groups do not differentiate between radiating matter, radiation, and irradiated matter \cite{Colclough.2011, Henriksen.1996, Lijnse.1990, Prather.2005}. The current state of research on pre-service teachers is precarious, apart from the study by Colclough \emph{et al.} \cite{Colclough.2011}. This lack of research clearly shows the need for an exploratory orientation to the following research question.

	\begin{quote}
		\textbf{RQ1}: What conceptions do pre-service teachers have in the topic area of radioactivity?
	\end{quote}

	\noindent In addition to the general description of conceptions about radioactivity from previous studies, it is questionable whether and how specific contextual references influence pre-service teachers' conceptions. Previous studies (\emph{e.g.} \cite{Eijkelhof.1990}) have shown that food irradiation is associated with an activation conception of the irradiated food. In addition to food irradiation, other application-related scenarios will be investigated and described. Building on our first research question, we also want to investigate what underlying structures in the sense of coordination classes emerge over various contexts. Part of investigating these underlying structures is also to observe possible context-dependencies.
	
	\begin{quote}
		\textbf{RQ2}: Which structures and context-dependencies occur in pre-service teachers' conceptions regarding radioactivity? 
	\end{quote}
	
	The context-dependency and structure of these conceptions allow for an interpretation of the material from the point of view of conceptual change. We investigate how pre-service teachers' conceptions are structured in the light of ``conceptual ecology'', and briefly include other points of view such as ontological reasoning. The description of half-life (\emph{e.g.} \cite{Hull.2021}) using the perspective of conceptual change is already well-advanced. Our broad approach aims at investigating more unknown aspects of ``conceptual ecology'', which may help to describe conceptions of radioactivity in the context of conceptual change in the future. In this way, the analysis of the collected material will not only describe the research gap from a descriptive perspective but also aims to provide scaffolding for further research.

	\section{METHOD}
	
	\subsection{Qualitative content analysis of semi-structured interviews}

	We conducted a semi-structured, problem-centered interview study to answer these research questions. To do this, we used four different prompts addressing the penetrating ability of ionizing radiation, food irradiation, radiopharmaceuticals, and nuclear power. We transcribed the interviews following the content-oriented semantic orientation of Dresing and Pehl with slight modifications \cite{Dresing.2015}. The pre-service teachers' expressions and content are most important for the interpretation. This led to the decision that phonological structures were left out in the transcription. The formation of the categories followed an inductive, or data-driven, approach. Therefore, the development of the coding system was not \emph{a priori} based on the physical concepts or theories of conceptual change but rather along the expressions of the pre-service teachers.
	
	We used semi-structured individual interviews to collect concepts. We conducted the interviews online using a video conferencing program, and analyzed them by applying qualitative content analysis. This analysis summarizes the raw material so that recurring ideas in the interviews are captured \cite{Mayring.2014}. We paraphrased the interviews and formed categories alongside the paraphrases. In light of conceptual change, categories or codes often refer to concepts used to describe the prompts. Before the study presented here, we developed a preliminary category system using three interviews. The interviews were not included in this work but were used to create the interview guide and category system. We reviewed the final category system for quality assurance, and any differences that arose were discussed and resolved. When the ``grounded theory'' indicated a transparent category system or reached theoretical saturation, additional interviews were deemed unnecessary and therefore waived \cite[cf.][]{Glaser.1967}.
	
	\subsection{Participants} \label{Particpants}
	
	To address our research questions, we invited pre-service teachers ($N=13$) who were enrolled in a teacher education program and had successfully finished a combined course on nuclear and atomic physics. Eight of the thirteen pre-service teachers studied physics combined with another STEM subject. Two of these interviews were conducted with pre-service teachers from another university. The aim was to exclude site-specific problems and to verify both the coding system and the theoretical saturation. The names of the pre-service teachers were assigned alphabetically for pseudonymization (\emph{e.g.} Georg appears in the tables as G).

	Participants were recruited during teaching sessions and by e-mail. Apart from passing the nuclear physics lecture at their respective universities, there were no further requirements. The selection process requirements meant that the sample was not randomly selected. Because the sample was limited to two universities, it is not representative with respect to the total population of pre-service teachers in Germany. The interviewer was not the lecturer of the nuclear physics course.
	
	The standard length of teacher-training programs in Germany is ten semesters, requiring teachers to be proficient in two subjects. The mean age of the participants was 24.3 years, and they had been enrolled in the study of physics education for a mean of 8.3 semesters in the study of physics for a teaching degree. Compared to the expected age of the participants, the average is slightly higher, which we attribute to course changes or similar. Seventy percent of the participants identified as male, while thirty percent identified as female. 
	
	\subsection{Interview guide}
	
	The interview guide we developed consists of four problem-centered prompts and an introductory phase (see Appendix A). The introductory phase established the students' definitions for radioactivity and their use of related quantities concerning nuclear physics. The prompts were introduced by sharing a picture or news article. The prompts describe different use cases of ionizing radiation or radioactive matter and represent different situations. Prompts 1 and 4 focus on nuclear processes and ionizing radiation, whereas 2 and 3 focus on the biological impact of ionizing radiation. 
	
	The interviews were started by asking how students would explain radioactivity as a term to fellow students who are not part of the physics program. We follow the pre-service teachers' use of terms (\emph{e.g.} ``radioactive'' radiation). The introductory phase covered radioactivity, radioactive materials, ionizing radiation, and half-life.
	
	\subsubsection{Prompt: Penetrating ability of ionizing radiation}
	
\begin{figure}[htp]
		\centering
		\includegraphics[width=0.33\textwidth, angle=0]{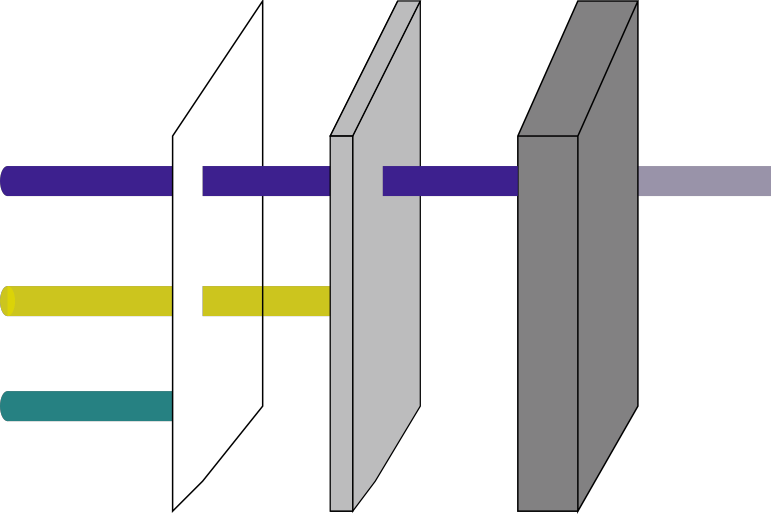}
		\caption{Representation of the penetrating abilities of different ionizing radiation through paper, aluminum, and lead. This prompt aimed to study conceptions related to the interaction of ionizing radiation with matter.}
		\label{img:penetrating}
	\end{figure}

	We assumed pre-service teachers would refer to three types of radiation (alpha, beta, and gamma) when describing radioactive processes. Therefore, our first prompt utilized a common way of representing these different forms of radiation and their respective penetrating abilities with paper, aluminum, and lead (see Fig.\ref{img:penetrating}). We further assumed that pre-service teachers attribute the difference in penetration ability to the size of the radiation particles. We drew this from the preliminary interviews that informed the development of the interview guide or coding system. These three types of radiation represent the types found in school and often in textbooks. In accordance with relevant literature, we also assumed that students might attribute the penetration ability to the energy content of the respective radiation \cite{Colclough.2011}. 
	
	We also assumed that the descriptions given by pre-service teachers are often associated with the risks regarding the three shown forms of radiation. In principle, this concept about the hazards posed by the different types of radiation is correct since alpha radiation, for example, cannot penetrate the skin. However, this does not apply to the incorporation of alpha emitters. In addition, it is important to note that different types of radiation often occur together within a material and that, for example, in the alpha decay of Am-241, one must also consider the gamma radiation released.
	
	The effect of ionizing radiation varies based on its energy. However, it is not solely determined by energy content, as different types of radiation have unique interaction mechanisms. This indicates that the differences between radiation types cannot be explained by energy alone. For example, while Compton scattering is the central process for the loss of energy for gamma radiation released by radioactive decays, the energy loss for charged particles of alpha and beta radiation is due to electromagnetic fields of nuclei according to the Bethe formula.
	
	\subsubsection{Prompt: Irradiation of food}

\begin{figure}[htp]
		\centering
		\includegraphics[width=0.33\textwidth, angle=0]{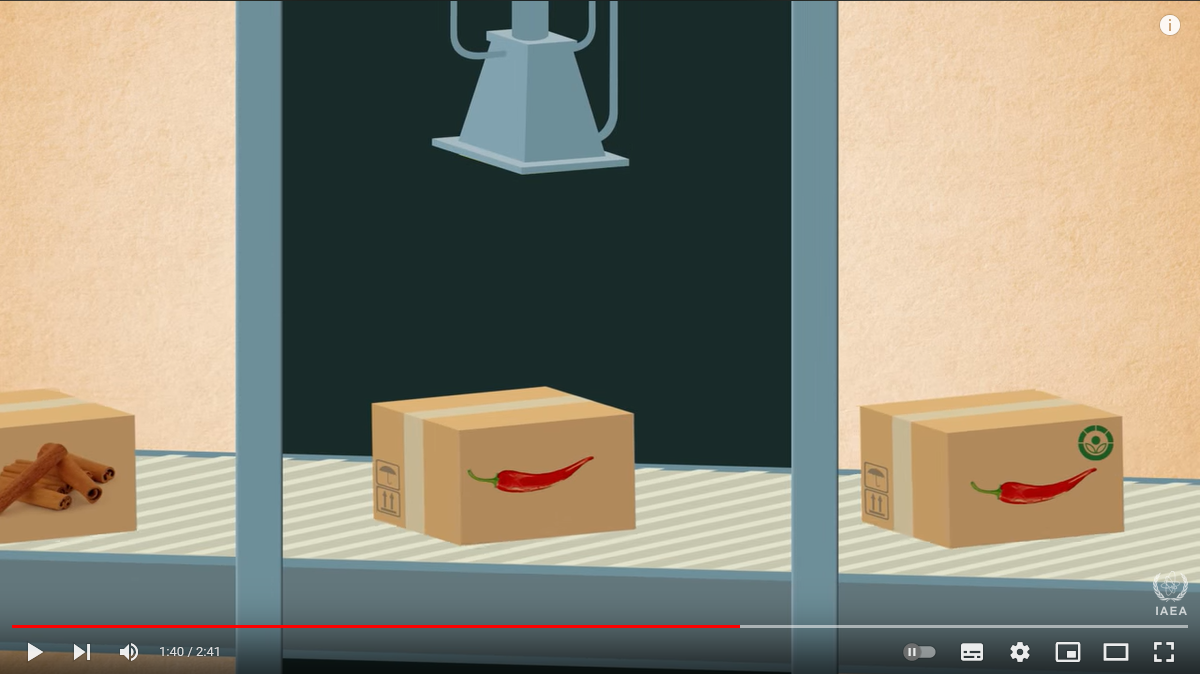}
		\caption{Frame of an educational video about the irradiation of food \cite{IAEA.2015}. This is comparable to the irradiation of strawberries discussed by Eijkelhof \cite{Eijkelhof.1990} (``Using Nuclear Science in Food Irradiation'', © IAEA, 2015).}
		\label{img:food}
	\end{figure}

	Food irradiation is a standard hygiene procedure, although its use in Europe is mainly limited to spices. Nevertheless, the reference to food opens up the possibility to distinguish between contamination and establishes a direct real-life connection for both students and pre-service teachers (see Fig. \ref{img:food}).
	
	\noindent Considering prompt 2, we asked pre-service teachers to explain the principles of food irradiation and evaluate them regarding its safety. This prompt provides an opportunity to address the conception of radiation storage. Storage can be compared with activation due to neutron radiation, although it does not occur for ordinary ionizing radiation. Here, we also wanted to investigate whether pre-service teachers differentiate between the terms \emph{radioactive matter}, \emph{ionizing radiation}, and \emph{irradiated object}. This prompt is distinct from the later prompts, which discuss radiopharmaceuticals and the Chernobyl accident, because it does not handle the dispersal of radioactive matter.
	
	Describing radiation storage or activation through food irradiation is a common approach in research on student perceptions \cite{Eijkelhof.1990}. We aimed to determine whether we can observe this conception within pre-service teachers. The prompt (see Fig. \ref{img:food}) presented seeks to replicate the findings from Eijkelhof \cite{Eijkelhof.1990} concerning the storage of ionizing radiation and the spread of radioactivity. Irradiation of food sterilizes it by destroying bacteria or other forms of life. Ionizing radiation leads to the formation of radicals via the ionization of matter, which in turn change or destroy DNA. Despite common belief, the food itself does not become radioactive.

	\subsubsection{Prompt: Radiopharmaceuticals}

\begin{figure}[htp]
		\centering
		\includegraphics[width=0.33\textwidth, angle=0]{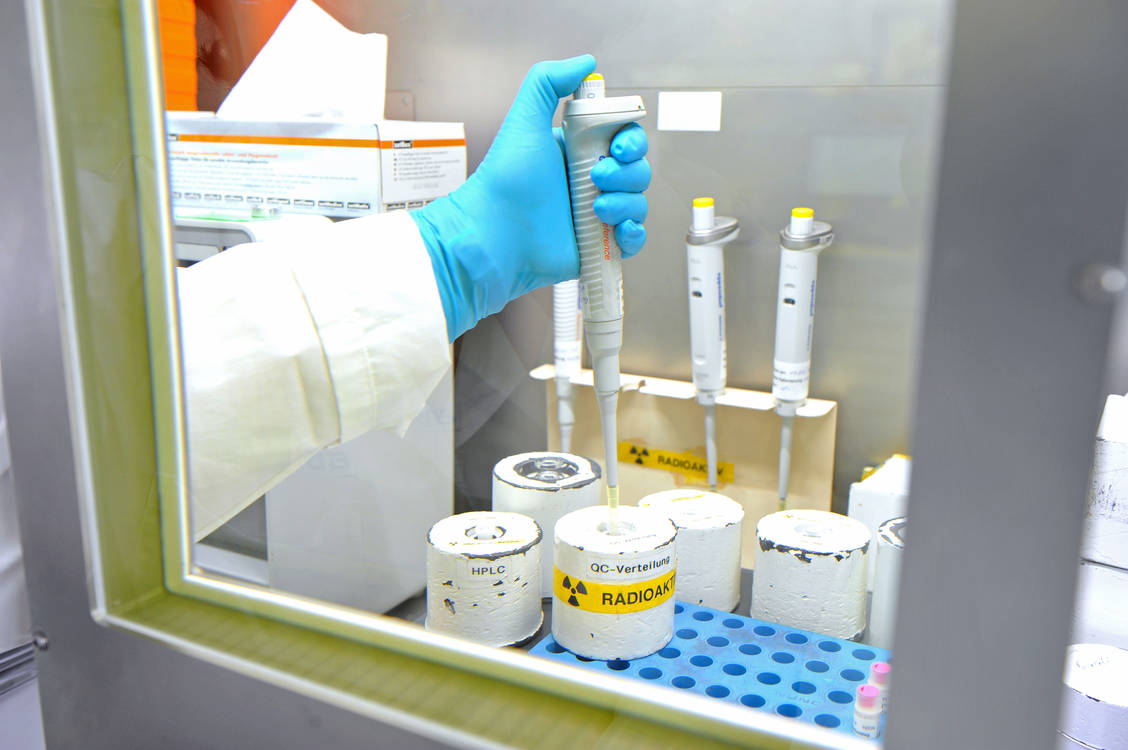}
		\caption{We introduced radiopharmaceuticals into the interview using this example picture. The aim here was to explain the mode of action of radiopharmaceuticals on a biological level. (HZDR/F.Bierstedt).}
		\label{img:radio}
	\end{figure}

	Radiopharmaceuticals, along with their desired and undesired effects on humans, require interdisciplinary explanation. 
	Radiopharmaceuticals and radiation therapy, which is similar on a biochemical level to the intake of radiopharmaceuticals, represent typical medical applications of radionuclides (see Fig. \ref{img:radio}). In the case of radiopharmaceuticals, the incorporation of radioactive matter must be addressed; in the case of radiation therapy (as with prompt 2), it must not be.

	When using radiopharmaceuticals, the goal is to attack cancer cells while avoiding healthy cells. Modern therapies might use local applications. One example of this is in the treatment of prostate carcinoma, which uses prostate antigens linked to a radionuclide. The strong effect on cancerous cells is due their increased cell division rate compared to healthy cells and the consequential impact of ionizing radiation on DNA while multiplying \cite{Steel.1996}. Radicals form due to the interaction ionizing radiation with molecules, which, in the case of the cell, are primarily water molecules. The radicals formed can attack nucleobases such as cytosine through oxidations and trigger a point mutation via a cascade of subsequent reactions. For example, the one-electron oxidation gradually converts cytosine to uracil, which, after methylation, is converted to thymine, which causes a point mutation \cite{Wagner.2010}.

	\subsubsection{Prompt: Radioactive exposure of wild boars}
	
	On April 26, 1986, a severe accident occurred at the Chernobyl nuclear power plant, transporting radioactive material by the wind to Central Europe and Germany. The disaster in Chernobyl challenges European countries today due to the radioactive isotopes transported over the continent. Since this event, various isotopes (such as Cs-137) and their biological interactions have become part of public knowledge. Due to its public importance, especially in Europe, we assumed that pre-service teachers in Germany would be aware of this event.
	
	In order to explain the circumstances surrounding the Chernboyl accident, it is first necessary for teachers to explain how nuclear power plants work and to describe nuclear fission as a physical phenomenon distinct from other nuclear processes. The release of energy by fission is crucial to this description, so one goal was to investigate whether pre-service teachers can distinguish nuclear fission from nuclear decay and to what they attribute the release of energy in nuclear power plants.
	
	Prompt 4 (a podcast description titled ``Radiant wild boars'') illustrates that to this day, wild boars contain radioactive Cs-137, which leads to one out of five wild boars being unfit for human consumption in southern Germany: 
	
	\begin{quote}
		``It has been 33 years since the disaster in Chernobyl, but the effects of the reactor accident are still measurable in this country. In some regions of southern Germany, every fifth wild boar is polluted with radioactivity. Their meat may not be sold.'' \cite{Urban.2019}
	\end{quote}

	\noindent Considering this prompt, we asked pre-service teachers to explain the temporal and factual links between the ``radioactive exposure'' of wild boars in southern Germany and the nuclear accident. This prompt provided opportunities to address the conception of the storage of radiation in materials in a different context than prompt 2 (irradiation of food) and to differentiate between the terms \emph{radioactive matter}, \emph{ionizing radiation}, and the \emph{irradiated object} using the measurable effects of Cs-137 on living things in southern German forests as a context-dependent scenario. The fire at the nuclear power plant in Chernobyl spread radioactive material across Europe, which then moved into the soil by precipitation. Isotopes such as Cs-137 are still detectable today due to their half-life, whereas this is not the case for I-131 due to its short half-life. Therefore, it is necessary to deal with the transport of radioactive material for proper clarification.

	In conjunction with this prompt, the pre-service teachers were asked to hypothesize why spent nuclear fuel could be more or less active compared to fresh nuclear fuel. Nuclear power plants draw their energy from the nuclear fission of fissile materials (neutron-induced fission). Compared to the energy that could be drawn from decay, the amount of energy gained from nuclear fission is about forty times larger. While the daughter nuclei are predictable during decay, this is not true for nuclear fission, which results in the creation of many different products. Decay and neutron-induced fission also differ in their reaction kinetics due to the necessity of neutron flux. Due to the relative neutron abundance of the fission products, they are typically beta-emitters. In addition to the actual fission reaction, neutron absorptions also occur, forming transuranium elements from the fuel. Compared to the enriched fuel (\emph{e.g.} U-235 or U-238), the half-lives of transuranium elements are significantly shorter (\emph{e.g.} Pu-239), accompanied by increased activity. Therefore, spent nuclear fuel has a higher activity than fresh fuel by magnitudes.

	\section{The case of Georg} \label{The case of Georg}
	
	At this point, we would like to discuss the case of Georg to provide an understanding of how we conducted the interview and what the expected answers were. We chose this case because it is representative of the general interview process. Georg (25 years old, identifies as male, enrolled in the teaching program for eight semesters) is studying a non-science subject besides physics. The names of Georg and other participants were pseudonymized as described in section \ref{Particpants}.

	\subsection{Case description}
	
	Georg describes radioactivity as the emission of radiation due to atomic processes. He also differentiates the emitted radiation using the concepts of particle and wave. Georg interprets the decay as the release of particles from the atomic nucleus, referring to representations of a sphere-particle model of the nucleus. He describes the three different radioactive decay processes as releasing different particles. The alpha, beta, and gamma radiation particles Georg describes are protons plus neutrons, electrons, or photons.

	As an example of radioactive materials, he cites uranium and its use in atomic bombs. He attributes the condition for nuclear decay to the size of the nucleus since it is is no longer stable above a specific limiting size:
	
	\begin{quote}
		
		Interviewer:\emph{ Do you know some materials that could be radioactive?}
		
		G:\emph{ The classic is uranium. We all know it from the atomic bomb. But in principle, everything above a certain [...], what's it called, certain size is simply / Because from then on, the nucleus is no longer stable and decays over time. I don't know now which number it is, it should be around 90, I would have guessed. [...]} (Georg, pos. 8-9)
	\end{quote}

	\noindent He identifies everyday radioactive materials by using the existence of certain isotopes and their radioactivity. For example, Georg refers to using radioactive carbon in date determination and radioactive iodine in medicine.

	Georg describes the ``strength'' of radioactivity as the number of measurable particles hitting a detector. The imprecise term ``strength'' was introduced by the interviewer. Next, he introduced the half-life of radioactive materials by their macroscopic interpretation while also describing the quantum mechanical statistical interpretation of the half-life for a single atomic nucleus as an average value for the decay time:
	\begin{quote}
		
		Interviewer:\emph{ [...] How could we now describe the frequency with which radioactivity occurs, with which something decays or so, for one kilogram for this or that substance?}
		
		G:\emph{ Ok. [...] So there is the so-called half-life. This is the average time unit until an atomic nucleus decays and releases radioactive particles. So a radiation particle is released. And this could be extrapolated to a mole, from the mole to the kilogram. Then one could indicate a frequency [...], yes, the frequency, how often such a particle theoretically should fly out of this kilogram. [...] }
		(Georg, pos. 16-17)
	\end{quote}

	\noindent The radiation released during radioactivity occurs in small concentrations in everyday human life, and the properties of the particular type of radiation are of interest for its shielding from his point of view. He describes the radiation released during radioactivity with the term radioactive radiation. Radioactive radiation, he says, often leads to the student's notion of activation when it interacts with matter. He separates this from contamination, in which radioactive matter is dispersed. He does not exclude an interaction of the radioactive radiation with matter because of its energy content. 
	
	\subsubsection{Penetrating ability of ionizing radiation}
	
	He introduces the importance of interaction in the case of gamma radiation by comparison of the absorption or emission of visible radiation. He attributes their low interaction to the lack of suitable levels of the atom:
	
	\begin{quote}
		
		Interviewer:\emph{ You have now talked about the fact that [particles of ionizing radiation] then find something to interact with. Why would that happen better with one kind than with the other?}

		G:\emph{ [...] Well, with gamma radiation, which are photons, i.e. light particles, it is clear in this respect that they can only interact if the energies are the same. [...] And since gamma radiation is a very high-energy radiation, I can well imagine that there is simply little that can do anything at all with this much energy, without then immediately becoming a free radical. [...] Why the alpha radiation is absorbed so well, [...] but because I have said before that it is a hydrogen nucleus, that is a very simple atomic nucleus in principle and the paper consists of atoms with atomic nuclei, which are heavier and larger than the hydrogen nucleus and I could imagine now simply that if one assumes now from the law of conservation of momentum, that thereby very quickly very much is absorbed. [...]} (Georg, Pos. 36-37)
	\end{quote}
	
	\noindent He imagines the interaction of alpha radiation, which he has thus far described as involving protons, ad-hoc as a mechanistic interaction (including a reference to conservation of momentum law). Similarly, beta radiation interacts with the electrons of the absorber. The danger for humans cannot be explained by their shielding ability alone, whereby he separates here again between contamination and irradiation. It is also apparent that the size of the particle plays a vital role for him.

	\subsubsection{Irradiation of food}
	According to Georg, damage to biological material is caused by destroying genetic material, which limits the divisibility of cells. The mechanism by which ionizing radiation does this is through the transfer of its energy to the desired structure. In the case of food irradiation, the structure in question is the microbe.

	\begin{quote}
		G:\emph{ So it's just that this radioactive radiation as such is a very high-energy radiation. That means, as soon as it interacts with something, it can release its energy. Highly energetic usually means that something gets broken. And what makes radioactive radiation dangerous for humans is that it can also destroy genetic material.} (George, pos. 29)
	\end{quote}

	In food irradiation, apart from the possibility of accidents during operation, he sees no danger to the consumer. He also notes that, the usability of the procedure could be subject to other unknown influencing factors, but he expressed these ideas without further evaluation of the safety of the procedure.

	\subsubsection{Radiopharmaceuticals}
	Georg describes the use of radiopharmaceuticals using the example of radioactive iodine for thyroid treatments. Radioactive iodine is, as he explains, preferentially absorbed by the thyroid gland, then reduces in size, and represents a minimally invasive treatment method. In addition to the thyroid gland, he talks about the use of radioactive radiation in the treatment of cancer: 
	
	\begin{quote}
		
		G:\emph{ [The radioactive radiation] would primarily kill the cells. Because I said that the radioactive radiation interacts with the genetic material, but of course it also interacts with cell walls. And if the cell wall is broken, the cell is dead. It's always a cost-benefit situation, you always have to think very carefully about how you use [radiopharmaceuticals], because the problem is that, as far as I know, radioactivity can also cause cancer. You just have to see how you dose it so that something intelligent comes out of it.} (Georg, pos. 63)
	\end{quote} 
	
	\noindent In doing so, he states that he used to think of this therapy as laser-like, with the radioactive radiation targeting the defective genetic material of the cancer cells in a controlled manner while estimating the side effects. He sees the biological impact of radioactive radiation as destroying cells or cell walls.

	\subsubsection{Nuclear energy and radioactive exposure of wild boars}
	
	Concerning the use of nuclear power, the conversion of mass into energy is of great importance, which he explicitly traces back to the mass-energy equivalence:
	
	\begin{quote}
		
		Interviewer:\emph{ [...] How do you imagine a nuclear power plant functions?}
		
		G:\emph{ [...] Well, as I understood it from my school days, it's basically like this: I use the radioactive decay as the source for simply making water warm, which I then run through a turbine, which then drives a generator, which in turn makes electricity. And yes, and what is radioactive about it is basically the kettle. [...] }

		Interviewer:\emph{ [...] How exactly does our kettle get warm? [...]}

		G:\emph{ [...] I have to think about it again. So [...] The joke is, as far as I remember, that in the splitting of the nuclei, mass is actually converted into energy with the famous Einstein formula E equals m c squared. This energy is, so energy is warm ((laughs)). [...]} (Georg, pos. 66-69)
	\end{quote}
	
	\noindent When asked to describe the term nuclear fission, he inaccurately conflates it with decay. During this process, the nucleus releases two equal parts. He attributes the release of energy in fission to the passage through the isotopes or the existence of decay chains:
	
	\begin{quote}
		
		Interviewer:\emph{ [...] Could [you explain] the term nuclear fission again?}

		G:\emph{ [...] Yes, so if I have it right, it's like this: I have an atomic nucleus. It has a certain number of protons and neutrons. And one speaks now of decaying. That doesn't mean that it decays into twenty million pieces, but it splits into two more or less identical parts, whereby a part is still given off in the form of radiation. Now I don't know at all whether this must always be a hydrogen nucleus then. Whether it must be also a positron or something. Because there was also beta plus and beta minus decay. [...] But now we come back to the pictures from the books. Normally it is just shown like this: I have a nucleus and it decays into two nuclei that look pretty much the same. If I remember it correctly, it is also like this, that you go through the isotopes. So the element as such does not change directly, but the isotopes change first. If you have gone through this often enough, then I think you get to the isotope of the next element. [...]} (Georg, pos. 70-71)
	\end{quote} 
	
	\noindent Later he states that especially two releasing particles in a chain reaction should excite the decay. He believes that these particles are controlled by lead rods: 
	
	\begin{quote}
		
		Interviewer:\emph{ [...] What do you think could be used to control this process of nuclear fission?}

		G:\emph{ Well, as far as I remember it, it's a chain reaction, so every split nucleus gives off two more splitting particles. I think that was the mistake earlier, we don't get two nuclear particles out of it of equal size, we get two particles out per nucleus mainly. [...] In any case, each split particle ensures, so to speak, that another two are split and this is then potentiated. And this can be controlled by absorbing these splitting particles, in German, this radioactive radiation. And this is done in the nuclear power plant, I think, in such a way that one can drive lead rods out and in. And this then controls how much radiation can actually have a splitting effect and how much is swallowed by the lead. [...]} (Georg, pos. 74-75)
	\end{quote}
	
	\noindent This is a non-normative answer to the prompt. Pre-service teachers typically fall into two categories: either the provision of energy is due to decay or fission. We coded this in Table V in column ``G'', as he used the concepts of decay and fission syncretically.

	Georg reasons that the activity of fresh fuel should be higher than that of used fuel, since the material can still be split here, and, as a result, usable energy is still present. He cannot give possible reasons for a higher activity of the fuel after its use. He attributes the hazards associated with using nuclear energy to control failures or vulnerabilities in the design and describes the question of final storage as problematic due to the long periods needed for its safe storage.

	He sees the presence of radioactively contaminated wild boars in Germany as an impact of the radioactive substances that leaked due to the Chernobyl accident. In response to Prompt 5, Georg responded with:
	
	\begin{quote}
		G:\emph{[The accident] has blown quite a lot of radioactive material into the atmosphere, which then turns out to be a dust cloud, so to speak[...]. Because the [dust], for example, then collects in certain places. [...] I don't know why, but I think I have read that mushrooms absorb radioactive material and store it in themselves [...].} (Georg, pos. 81)
	\end{quote}
	
	\noindent This is the normative answer to the prompt, in which pre-service teachers ascribe the presence of radioactively contaminated wild boars to the distribution of radioactive substances that leaked due to the Chernobyl accident. We coded this in Table II as a red box under ``G'' (see Table II), as he modeled ionizing radiation as transported material. While this distinction appears in many interviews, it does not appear in all of them. As will be seen later regarding the interview of Lars, pre-service teachers' ``radioactive contamination'' of wild boars can also be attributed to the spread of radioactive radiation.

	\subsection{Case analysis}
	
	Following the school-typical separation of particle and wave radiation, Georg separates the three types of radiation into two classes, with references to known representations. Georg's reasoning does not clearly distinguish between wave and particles, as the photons are considered gamma radiation particles. The released particles coordinate the mechanisms over which the radioactive radiation can interact with matter. It becomes clear that quantum mechanical interactions do not play a prominent role for Georg in describing radioactive radiation. He describes alpha and beta radiation phenomenologically and their interactions mechanistically. We expected this because of the semi-classical approach to many nuclear physical phenomena. Only for gamma radiation does he use a quantum mechanical argument that considers energy levels. It is also noteworthy that he mentions a possible ``interaction if energies are the same''. This statement suggests that Georg may consider the energy levels of particles involved in a radioactive process as a factor influencing their interactions.

	Georg classifies radioactive substances along the periodic table; additionally, the isotopes of certain elements become the radioactive version of the element (\emph{e.g.} radioactive iodine). The size and associated instability of the nucleus become central features of radioactive substances. However, this rigid view is diluted by radioactive variants, where no controlling effects like those of the nucleus size are recognizable. The orientation to the periodic table of the elements plays a significant role for Georg. Radioactivity as a physical phenomenon is described imprecisely by the terms strength and frequency by the terms activity and measurable count rate. There is no separation concerning energy, which is the gold standard to distinguish between isotopes experimentally. To disregard energy at this point makes the description of different isotopes more difficult. The description thus remains on a phenomenological level and is not able to reveal concrete interaction mechanisms.

	Following the analogy of optical radiation, he attempts to describe energy absorption within the cells. The separation between irradiation and contamination is factually appropriate regarding previous teaching experience, and his description of radioactive contamination of wild boars, ingestion of radiopharmaceuticals, and general radiation protection instructions reflect this.

	To describe the energy provision in nuclear power plants he uses both concepts of decay and fission. In interpreting Georg's description, a hybridized model of fission and decay emerges. In this model, the ideas of decay and fission are conflated, which can be seen, in the use of the decay series. The use of technical language follows a syncretic pattern. Although Georg does not correctly separate the concepts of fission and decay, the use of decay is superficially sufficient to explain the release of energy in a nuclear power plant. Thus, we expect that Georg would be unable to correctly distinguish between a nuclide battery and a nuclear power plant.

	The interviews suggest that Georg may draw analogies to familiar concepts to understand the central concept of energy better. He compares a nuclear power plant to a kettle, indicating similarities in how energy is generated or released. This analogy serves as a mental framework for Georg to grasp the concept of energy in the context of radioactivity. The idea of mass equivalence is also relevant to Georg's considerations. This understanding could influence his reasoning about the behavior of particles during radioactive processes, including fission and decay. He reduces the functioning of nuclear power plants to the question of energy release alone, and the concrete processes become blurred. Additionally, the interviews mention that fission and decay are undifferentiated (\emph{e.g.} by regulating the energy output with lead rods), implying that Georg may see similarities or connections between these two processes. This may indicate that Georg considers energy as a critical factor in both fission and decay and that he does not distinguish between them in terms of energy considerations.

	Georg's comprehension of radioactivity heavily relies on the central concept of energy, which plays a pivotal role in his reasoning and considerations. This concept significantly impacts how he perceives particle behavior and interactions and shapes his analogical reasoning, ideas about mass equivalence, and categorization of various radioactive processes. Considerations concerning energy also occur with other students (see \ref{RESULTS AND ANALYSIS}) and are also normatively considered a central aspect of radioactivity or ionizing radiation description.

	\section{RESULTS AND ANALYSIS} \label{RESULTS AND ANALYSIS}
	
	We used binary-coded tables in the following representations of the codes present in the interviews (\emph{e.g.} Table \ref{table:Model of ionizing radiation}). For example, the presence of code K-5 (``decay/radioactivity'') describes that pre-service teachers attribute the provision of energy to the decay or radioactivity of uranium.

	The following results, or the associated categories, are separated into two types. The second categories are those referred to below as ``specific'' categories. Based on these categories, we identified conceptions. Specific categories are characterized either by having contradictory counterparts (nuclear power plants draw their energy from decay vs. fission, K-5 and K-6) or by being differentiated by their depth of reference (B1 to B-3 for the point of action from a biological point of view). Specific categories are of particular importance for descriptions of conceptions (\emph{e.g.} K-5, Nuclear power plants use the decay of uranium to provide energy). A quantified
	representation is not possible in this work due to its semi-structured approach.

	The categories referred to below as ``unspecific'' represent indicative characteristics of radioactivity that the pre-service teachers cited but to which we did not attribute conceptual properties. In the Appendix, we marked these codes with a star. These include statements such as the instability of the nucleus as a necessity of radioactive processes that all students cited (\emph{e.g.} R-1*, Instability). This category classification partly results from how the interview proceeded since we did not observe conflicting answers in the associated replies.

	The description of the coding system, including anchor examples, is provided in Appendix B. We included a summary of the occurrence of all codes in Appendix C. In the following tables, we marked the presence of a code in an interview with differently marked fields. A checkmark indicates the presence of a code in an interview and an X indicates its absence. We will now show which ideas are associated with our research questions depending on the respective scenarios we presented.

	\subsection{Properties of radioactivity}
	
	The findings presented in this section represent unspecific results (R-1* to R-17*, see Appendix B), which are listed to show how pre-service teachers understand radioactivity.
	
	The condition for the occurrence of radioactivity is the instability of the atomic nucleus (R-1*). The characteristic of radioactivity is the nuclear decay under the release of particles or the emission of ionizing radiation (R-2* and R-3*). The transformation of a nucleus from one element into another, or the transmutation that occurs in decays, is a less-described feature of radioactivity (R-4*). Less than half of the students describe ionization as a central property of ionizing radiation released during radioactivity (R-5*). For the case of the interaction of ionizing radiation with non-living matter, some students consider it possible (R-6*), but only a few pre-service teachers named concrete processes.

	Pre-service teachers describe radioactive materials in terms of the elements, usually elements with an atomic number greater than 84, particularly plutonium and uranium. Sometimes, when they refer to radioactive variants, they also mention elements with a smaller atomic number (R-7*). They associate radioactivity with ionizing or ``radioactive'' radiation and radioactivity occurs for them naturally (R-8*, R-9*). 
	
	They reduce ionizing radiation associated with radioactivity to three major varieties: alpha, beta, and gamma radiation (R-10*). Alpha radiation is, to them, the emission of helium nuclei, beta radiation is the emission of electrons, and gamma radiation is the emission of electromagnetic waves or photons (R-11* to R-13*).

	The pre-service teachers describe the ``strength'' of radioactivity by the activity, i.e., the number of decays per unit of time (R-15*). The few pre-service teachers who describe the ``strength'' of radioactivity in terms of the energy of the emitted radiation are coded with R-14* (R-14*). Next, they recount the time behavior of radioactive processes using half-life. The description on the macroscopic level is carried out appropriately by a decrease in the number of radioactive nuclei of the substance. Occasionally, they refer to a decrease in the mass of the radioactive substance (R16*). Finally, a description of the microscopic statistical-quantum mechanical interpretation takes place in eleven of thirteen interviews. Here, pre-service teachers connect the half-life with a corresponding probability of the nucleus decaying within a given period (R-17*).

	\subsection{Models of ionizing radiation} \label{Models of ionizing radiation}
	
	\subsubsection{Results}
	
	Pre-service teachers attribute radioactive contamination (such as in wild boars) partly to the transport of radioactive material (M-1, see Table \ref{table:Model of ionizing radiation}). It is also partly attributed to the distribution of radiation, characterized by the formation of residues or the excitation of the object (M-3, see Table I). We excluded the activation due to neutron radiation or nuclear reactions. Less frequently, they attribute the ``spread'' of radioactivity to the transport of radioactivity itself (M-4, see Table \ref{table:Model of ionizing radiation}). Some pre-service teachers also explicitly rejected the activation due to ionizing radiation (M-2, see Table \ref{table:Model of ionizing radiation}). This rejection occurred without prompting it explicitly.

	\begin{table}[htp]	
		\centering
		\caption{\raggedright{How is ``radioactivity'' transported?} \linebreak \footnotesize{A checkmark indicates the presence of a code in an interview and an X indicates its absence.}}
		\label{table:Model of ionizing radiation}
		\begin{tabularx}{\linewidth}{ l XXXXXXXXXXXXX r}
			\toprule
			& A & B & C & D & E & F & G & H & I & J & K & L & M & $\Sigma$ \\
			\specialrule{0.5pt}{0.25pt}{0.25pt}
			\thead{Transport of material \\ (M-1) \hspace*{7em}}	& \checkmark & \checkmark & \checkmark & X & \checkmark & \checkmark & \checkmark & \checkmark & \checkmark & \checkmark & X & X & \checkmark & 10 \\
			\specialrule{0.5pt}{0.25pt}{0.25pt}
			\thead{Rejection of activation \\ (M-2) }&    X         & X & X & X & \checkmark & X & \checkmark &X  & X & \checkmark & \checkmark & \checkmark & \checkmark & 6 \\
			\specialrule{0.5pt}{0.25pt}{0.25pt}
			\thead{Activation \\ (M-3)} &  \checkmark  & \checkmark & \checkmark & \checkmark & X & \checkmark & X & X & \checkmark & \checkmark & X & \checkmark & X & 8 \\
			\specialrule{0.5pt}{0.25pt}{0.25pt}
			\thead{Transport of \\ radioactivity (M-4) } & X & X & X & \checkmark & X & X & X & X & \checkmark & \checkmark & X & X & X & 3 \\
			\specialrule{0.5pt}{0.25pt}{0.25pt} 
		\end{tabularx}
	\end{table}
	
	At this point, we did not take prompt 4 (Radioactive exposure of wild boars due to the Chernobyl incident) into account. We will be able to show that even if students explicitly reject radiation storage, it can occur at prompt 4. This exclusion leads to a finer image. While this gives us a perspective into the overall use of the conceptions throughout the interview, Table \ref{table:Model of ionizing radiation_used} shows which scenarios were associated with different models of ionizing radiation.

	\begin{table}[htp]	
		\centering
		\caption{\raggedright{Uses of different models of ionizing radiation.} \linebreak \footnotesize{A checkmark indicates the use of activation (M-3), an X indicates the use of transport of material (M-1), and a tilde indicates an undifferentiated mix of various conceptions (M-1, M-3, and M-4). Empty spaces indicate the absence of the corresponding codes.}}
		\label{table:Model of ionizing radiation_used}
		\begin{tabularx}{\linewidth}{ l XXXXXXXXXXXXX r}
			\toprule
			& A & B & C & D & E & F & G & H & I & J & K & L & M & \\
			\specialrule{0.5pt}{0.25pt}{0.25pt}
			\thead{Irradiation of food \\ (Prompt 2) \hspace*{7em}}	& \checkmark &  & \checkmark & \checkmark & & \checkmark & &  & \checkmark &  &  & \checkmark & & \\
			\specialrule{0.5pt}{0.25pt}{0.25pt}
			\thead{Wild boars \\ (Prompt 4) }&  X & X & X & $\sim$ & X & X & X &X  & $\sim$ & $\sim$ &  & X & X & \\
			\specialrule{0.5pt}{0.25pt}{0.25pt}	
		\end{tabularx}
		
	\end{table}

	Pre-service teachers use the activation model (M-3) when discussing food irradiation (as in prompt 2). ``Activation'' here means the storage of radiation. From a normative perspective, ``radiation'' is only such when it is being sent out from a radioactive body and, as such, cannot be stored either prior to or after this traveling. In the second prompt's discussion, pre-service teachers actively rejected this assumption (Rejection of activation, M-2, see Table I).

	In contrast to the model used to answer prompt 2, the vast majority of interviewees attributed the contamination of wild boars (prompt 4) to the transport of radioactive material (M-1). This switch is evident in four of thirteen pre-service teachers and detectable within limits in two other pre-service
	teachers (see Table II).

	\subsubsection{Interpretation} \label{Models,Interpretation}
	Pre-service teachers, in contrast to high school students, usually associate the transport of radioactivity with mass transport. Codes M-3 and M-4 identify the distribution of radiation and radioactivity and are nevertheless present in pre-service teachers. In addition, pre-service teachers suspected that food irradiation (Prompt 2) could potentially lead to further decays in the food:
	
	\begin{quote}
		\raggedright{A:\emph{So it is already residues, because by the irradiation one stimulates practically also further atomic nuclei to the decay. But. [..] It depends in any case on the type of radiation that it is then just harmless or more harmless than with chemistry, somehow.}} (Anja, pos. 59) \endgraf
	\end{quote} 
	
	\noindent The use of the concept of activation due to ionizing radiation occurs most prominently in prompt 2. Here we specifically ask about the expected negative and positive consequences of irradiation in the food industry. The students who explicitly exclude radiation storage (M-2) generally use a consistent description of the model of ionizing radiation. An explicit negation of this model allows the assumption that this is part of a learning process:
	
	\begin{quote}
		L:\emph{ What I noticed, or what is always such a common misconception, is yes, when things come into contact with radioactive radiation, many assume that themselves become a radiator.} (Lars, pos. 35)
	\end{quote} 
	
	\noindent It is noteworthy that even if students explicitly negated the storage of radiation, they sometimes switched back to an activation concept:

	\begin{quote}
		Interviewer:\emph{ Would you see any disadvantages in using such irradiation on food?}
		
		L:\emph{ It is always questionable what kind of high radiation dose they get. Because we always take it into our human body. They themselves do not decay, but the radioactive radiation is still detectable. And we have to be clear about this. What is healthy for humans when they absorb radiation? What of it would be degradable and when does it become dangerous to health? [...]} (Lars, pos. 48-49)
	\end{quote}
	
	\noindent In Lars' answer, it becomes clear that this idea nevertheless persists despite the explicit exclusion of the activation of the material. While he negates the storage of radiation in the beginning of his answer, this negation is abandoned by the following paragraphs. We observed this context-dependency in the previously cited case of Anja and other pre-service teachers (Table II).

	Assuming that the negation of storage is due to reflected learning of the pre-service teachers, it shows that this subgroup of students could already identify student conceptions concerning radioactivity as their own prior conceptions. Biological references tease out the notion of radiation storage, which we understand as falling back to the concept of storage in the context of food irradiation. Interestingly, food irradiation is a strong framing for using an activation concept to describe ionizing radiation. In contrast to this is the use of transport (M-1) for describing wild boars in southern Germany. Comparing those two prompts shows what impact framing or context-dependency can have on experienced pre-service teachers.

	When comparing the models, a distinction between matter and process is evident. When students confuse radioactive substances, radioactivity, and ionizing radiation, looking at the ontologies of substance and process can be helpful. Chi's ontological categories of substance and process provide a practical framework. Radioactivity is not a tangible substance but a property of specific nuclei, and comprehending the processes involved in radioactive transformations is essential. Students can understand this phenomenon precisely and comprehensively by clearly defining and differentiating between matter and process when teaching about radioactive substances, radioactivity, and ionizing radiation.
	
	Understanding the concept of radioactivity requires one to differentiate between matter and process. Radioactivity refers to the ability of a nucleus to undergo a transformation process rather than to being a tangible substance. It is a property of specific nuclei that involves the spontaneous emission of particles or energy to achieve a more stable state. Students must grasp this concept, as without it, confusion and misunderstandings about radioactivity can arise.

	\subsection{Penetrating ability of ionizing radiation}
	
	\subsubsection{Results}

	The penetrating ability of ionizing radiation allows students to estimate the risk associated with the respective type of radiation (D-1, see Appendix). However, they often add that this is no longer true if radioactive substances are incorporated into the body (D-2). They attribute the penetration ability to the size of the radiation (D-3, see Table III). With the concept of size, we summarize the concepts of mass and volume. In addition to the size, pre-service teachers use the energy content of the radiation to explain its penetrability. Gamma radiation is often said to have the largest energy (D-4). They mention the interaction of particles with matter to explain their penetrability (D-5), often without arguing with specific mechanisms. Sometimes the penetration capability is described as a function of the density of the absorbing material (D-6), which the student added without being prompted by a question from the interviewer.
	
	\begin{table}[htp]	
		\centering
		\caption{How do you explain the different behavior of the different types of ionizing radiation? \linebreak \footnotesize{A checkmark indicates the presence of a code in an interview and an X indicates its absence.}}
		\label{table:Penetrating ability of ionizing radiation}
		
		\begin{tabularx}{\linewidth}{ l XXXXXXXXXXXXX r}
			\toprule
			& A & B & C & D & E & F & G & H & I & J & K & L & M & $\Sigma$ \\
			\specialrule{0.5pt}{0.25pt}{0.25pt}
			\thead{Size \\ (D-3) \hspace*{7em}} & \checkmark & X  & X  & X  & \checkmark & \checkmark & \checkmark & \checkmark & \checkmark & X  & \checkmark & \checkmark & \checkmark & 9    \\
			\specialrule{0.3pt}{0.25pt}{0.25pt}
			\thead{Energy \\ (D-4)}  & X  & X  & \checkmark & \checkmark & X  & X & X  & \checkmark & \checkmark & \checkmark & \checkmark & X  & \checkmark & 7    \\
			\specialrule{0.3pt}{0.25pt}{0.25pt}
			\thead{Interaction \\ (D-5)} & \checkmark & \checkmark & X  & X  & \checkmark & X  & \checkmark & \checkmark & \checkmark & X  & \checkmark & \checkmark & X  & 8    \\
			\specialrule{0.3pt}{0.25pt}{0.25pt}
			\thead{Density of absorber \\ (D-6)} & X & X & X & X & X & \checkmark & \checkmark & X & X  & \checkmark & \checkmark & X  & X  & 4 \\
			\specialrule{0.3pt}{0.25pt}{0.25pt} 
		\end{tabularx}
	\end{table}

	\subsubsection{Interpretation}
	
	The assumption of hazards based on the penetrating power of the different types of radiation represents typical school knowledge. However, only eight out of thirteen pre-service teachers succeed in describing these hazards as a result of both the penetrating power of ionizing radiation as well as the incorporation into the body. Illustrations, as shown in prompt 1, make this simplification possible. If we consider, for example, the decay of Am241, it becomes clear that several types of radiation often emanate during nuclear decay. This is also obvious for the decay series.

	The concept of size (D-3) is crucial in describing the penetrating ability of ionizing radiation for pre-service teachers:
	
	\begin{quote}
		H:\emph{ By size. So with alpha-, beta- you can still talk about this ``classical particle'', in quotes, so to speak. So I have a helium nucleus, or an electron, which is emitted. And just by their extension, by their volume, [they] can be blocked then still relatively well by matter.} (Hannes, pos. 31) 
	\end{quote}
	
	\noindent The concept of size can be explained mechanically, just like a cannonball. In the case of pre-service teachers, they consider the volume or mass of an object, which is categorized as the size concept. The object's energy (D-4) is also considered to determine its penetration capability:
	
	\begin{quote}
		H:\emph{ [...] Maybe with the speed with which they escape. So, that means the alpha particles weigh relatively much. That means they are probably not accelerated as fast as a beta particle. So, if an alpha particle would be fast enough, it could certainly break through a piece of paper.} (Hannes, pos. 33)
	\end{quote}
	
	\noindent The description of the energy of the emitted radiation is relevant for the description of the effective cross-sections. However, the interaction mechanisms must be considered, which differ for the different types of radiation. Thus, using energy to describe the penetrating power is not inherently wrong but incomplete. If we consider the gamma radiation occurring in nuclear decay, the most frequent interaction process here is Compton scattering, which does depend on energy. Nonetheless, gamma photons with more energy are said to be more penetrating:

	\begin{quote}
		Interviewer:\emph{ [...] Why does [..] one type [of ioninizing radiation] come further than the other?}
		
		I:\emph{ The energy? Well, especially gamma rays exist in different energies, so I think that the ones with higher energy go further than the ones with lower energy. [...]}
		(Ina, pos. 36-37)
	\end{quote}
	
	\noindent Along with size and energy, interaction is also considered a factor. The interaction capability (D-5) refers to the presence of possible processes:
	
	\begin{quote}
		G:\emph{ [...] So shielding, as far as I understand it, always has something to do with interacting. [...] Would it be so that the alpha rays, which were, if I remember correctly, the hydrogen nucleus, thus the proton and the neutron, that they already find enough other [...] atoms in a sheet of paper with which they can interact, so that these then quasi take up this energy. Or this simply then absorb, reflect, scatter. [...]} (Georg, pos. 35)
	\end{quote}
	
	\noindent In the excerpt shown above, Georg discusses the interaction of individual particles. These particles must find a suitable partner, which he imagines is easier for alpha radiation. A loss of energy can characterize the interaction. At the end of his quotation, he also deals concretely with different interaction mechanisms. In doing so, he forms a mechanistic understanding of interactions, but the causal interactions remain unfamiliar. One can speak here of an archaic model of the effective cross-section. Regarding the depicted prompt, size is reinforced as the decisive quantity. Cross-sections of effects were the subject of lectures at both universities that the pre-service teachers had attended.

	The interaction of ionizing radiation with matter is a process in the sense of Chi. However, what is significant in describing the process are the factors involved. For pre-service teachers, these are size, energy, and interaction. The size of the radiation particles establishes a link to a geometric understanding of interaction, such as that used in the impact theory of chemical reactions. Likewise, one can underestimate that the interaction of ionizing radiation with matter typically involves processes that are no longer adequately described with a geometric understanding.

	Size, interaction, and especially energy are good candidates for coordination classes in the context of ionizing radiation and its interactions with matter. These lower-level physical entities can be integrated into higher-level concepts related to the behavior and effects of ionizing radiation. For example, pre-service teachers often use size and energy to describe the penetrating abilities of ionizing radiation, but size can be problematic when considering quantum objects. By understanding these concepts and their relationships to each other, students can develop a more comprehensive and flexible understanding of ionizing radiation and its effects. 
	
	When it comes to radioactivity or nuclear physics, many concepts related to energy can be overwhelming. Energy is a good candidate for a coordination class in nuclear processes because it is a fundamental concept in understanding the behavior and effects of radioactive decay. It can be broken down into different sub-classes, such as kinetic energy, potential energy, and nuclear energy, which are all relevant for understanding the different types of energy involved in nuclear processes. Understanding the different forms and amounts of energy involved in nuclear processes is crucial for predicting and mitigating their effects, such as radiation exposure and nuclear waste management. Therefore, energy is a crucial coordination class that can help students develop a more comprehensive and flexible understanding of nuclear processes.

	\subsection{Biological effects of ionizing radiation}
	
	\subsubsection{Results}
	Pre-service teachers describe the biological effect of ionizing radiation in all cases as damage to biological structures or general damage to the living being (B-1, see Table \ref{table:Biological effects of ionizing radiation}). We differentiated between biological structures, including all biological structures larger than or equal to a cell, and biochemical structures, whose description in radioactivity is aimed at their chemical bonds. All participants who identified as female referred to damage to reproductive organs; this was true for only a smaller proportion of participants who identified as male. This finding, while interesting, must be considered cautiously, considering the small sample size. Some pre-service teachers attribute this damage to biochemical changes (\emph{e.g.}, of the DNA or hereditary material), and the necessary processes are described mechanistically as collisions (B-2, see Table IV). In rare cases, they extended this by a stochastic description of the damage to the DNA, by which the occurrence of cancer is explained (B-3, see Table IV).

	\begin{table}[htp]	
		\centering
		\caption{Where do the biological effects of ionizing radiation take place? \linebreak \footnotesize{A checkmark indicates the presence of a code in an interview and an X indicates its absence.}}
		\label{table:Biological effects of ionizing radiation}
		
		\begin{tabularx}{\linewidth}{ l XXXXXXXXXXXXX r}
			\toprule
			& A & B & C & D & E & F & G & H & I & J & K & L & M & $\Sigma$ \\
			\specialrule{0.5pt}{0.25pt}{0.25pt}
			\thead{Biological structures \\ (B-1) \hspace*{7em}}& \checkmark & \checkmark & \checkmark & \checkmark & \checkmark & \checkmark & \checkmark & \checkmark & \checkmark & \checkmark & \checkmark & \checkmark & \checkmark & 13 \\
			\specialrule{0.3pt}{0.25pt}{0.25pt}
			\thead{Biochemical structures \\ (B-2)}&   \checkmark         & \checkmark & X & X & \checkmark & \checkmark & \checkmark &\checkmark  & \checkmark & X & \checkmark & \checkmark & \checkmark & 10 \\
			\specialrule{0.3pt}{0.25pt}{0.25pt}
			\thead{Stochastic effect \\ (B-3)} &  X  & \checkmark & X & X & X & X & \checkmark & \checkmark & X & X & X & \checkmark & X & 4 \\
			\specialrule{0.3pt}{0.25pt}{0.25pt} 
		\end{tabularx}
	\end{table}
	
	\subsubsection{Interpretation}

	The biological effects of ionizing radiation are known to students in terms of their carcinogenic properties. Given the forms of presentation of ionizing radiation in the media, this finding is in line with expectations. Students refer to directly observable phenomena that are conducive to illustrating the risks of ionizing radiation. In addition to the macroscopic interpretation (B-1), some students refer to hereditary information on a biochemical level, including a particle-level mechanism (B-2). A transfer of energy is central to the biological effect:
	
	\begin{quote}
		K:\emph{ I mean, bacteria are also organisms and when they are exposed to such a high energy, maybe you can understand it in a similar way as when bacteria are exposed to a high thermal energy. [...] I can imagine that it is simply too much for the bacteria and that they then die.} 
		(Konstantin, pos. 48-49)
	\end{quote}
	
	Table \ref{table:Biological effects of ionizing radiation} shows a hierarchical understanding of the biological effect of ionizing radiation. If pre-service teachers describe the stochastic effect of ionizing radiation, they can also explain the effect on biochemical and biological structures. However, if we consider the processes at hand here, we must speak of an emergent process in the case of the effect. Some pre-service teachers take a direct approach to understanding the biological effects of ionizing radiation, while others may take a more nuanced, stochastic approach. The direct approach would likely focus on the immediate physical impact of ionizing radiation on large-scale biological structures. In contrast, the stochastic approach would consider the probabilistic nature of radiation interactions with those structures. Pre-service teachers generally need to use a more comprehensive description that includes stochastic risks. These findings further highlight that students describe a simplified mechanism of action that is still sufficient to understand the hazard and safety measures. It is adequate because they know bacteria die when faced with ionizing radiation. However, it is insufficient because the actual processes that lead to the change of DNA and their stochastic nature are unknown. When looking at the effects of radiation on biological components, there are two approaches: the direct approach and the stochastic approach. The direct approach oversimplifies the interactions between radiation and larger biological structures, believing that radiation is the direct cause of harm. The stochastic approach takes into account the probabilistic nature of radiation interacting with small-scale biochemical structures and recognizes that the effects of radiation can vary greatly depending on factors such as exposure intensity and individual biological susceptibility. Furthermore, the stochastic approach recognizes that the physical impact of the radiation does not solely determine the biological effects of ionizing radiation but may also arise from complex interactions between the radiation and the biological system.
	
	Thus, from Chi's ontological perspective of emergent processes, the stochastic approach to understanding the biological effects of ionizing radiation is likely to provide a more comprehensive and accurate picture of these effects. Furthermore, this approach recognizes that the effects of ionizing radiation are not simply a direct result of physical damage to biological structures but instead arise from complex interactions between radiation and the biological system. By considering these interactions, the stochastic approach can provide a more nuanced understanding of the complexities connected to the biological impact of ionizing radiation.

	\subsection{Nuclear power plants}
	
	\subsubsection{Results}
	
	In addition to the possibility of accidents (K-4*, see Appendix B), the question of the final repository of nuclear waste is part of the description of the risks of operating nuclear power plants (K-3*, see Appendix B). Complementary to this, possible reductions in the emission of greenhouse gases as well as restrictions regarding the supply of ``raw'' resources, were named and summarized under the category ``socio-ecological aspects'' (K-1*, see Appendix B). Control mechanisms of nuclear power plants were also described (K-2*, see Appendix B).

	\begin{table}[htp]	
		\centering
		\caption{How do nuclear power plants get their energy? \linebreak \footnotesize{A checkmark indicates the presence of a code in an interview and an X indicates its absence.}}
		\label{table:How do nuclear power plants get their energy?}
		
		\begin{tabularx}{\linewidth}{ l XXXXXXXXXXXXX r}
			\toprule
			& A & B & C & D & E & F & G & H & I & J & K & L & M & $\Sigma$ \\
			\specialrule{0.5pt}{0.25pt}{0.25pt}
			\thead{Decay/radioactivity \\ (K-5) \hspace*{7em}}	& \checkmark & X & \checkmark & X & X & \checkmark & \checkmark & \checkmark & \checkmark & \checkmark & \checkmark & X & X & 8 \\
			\specialrule{0.3pt}{0.25pt}{0.25pt}
			\thead{Fission \\ (K-6) \hspace*{7em}}&  X  & \checkmark & X & \checkmark & \checkmark & \checkmark & \checkmark & X & X & X & X & \checkmark & \checkmark & 7 \\
			\specialrule{0.3pt}{0.25pt}{0.25pt}
			\thead{Mass defect \\ (K-7) }&   X         & X & X & X & X & X & \checkmark &\checkmark  & X & X & X & X & X & 2 \\	
			\specialrule{0.3pt}{0.25pt}{0.25pt}
		\end{tabularx}
	\end{table}
	
	Nuclear power plants are said to derive their energy from uranium decay or inherent radioactivity (K-5, see Table \ref{table:How do nuclear power plants get their energy?}). When describing a particular decay, it is common for pre-service teachers to refer to decay chains. However, there is an alternative perspective that suggests energy comes from nuclear fission (K-6). In certain situations, both viewpoints are relevant. This is demonstrated by referring to two nuclei of similar size along with the previously mentioned decays (K-5). In two cases, the energy was said to be gained from the various mass defects of the nuclei (K-7). This description is complementary to K-5 and K-6 in character.

	In all cases, the reason for the possible higher activity of fresh nuclear fuel is greater energy (K-8, see Table \ref{table:Activtity of fresh and spent nuclear fuel}). Higher activity of spent nuclear fuel is attributed to the multiplication of the number of nuclei by fission (K-11), activation by storage of radiation (K-10), or dependence on nuclear reactions of nuclear fuel not described in detail (K-9). The activation by storage of radiation does not refer to the absorption of neutrons.
	In such a case, we assigned this to the nuclear reactions category (K-9).
	
	\begin{table}[htp]	
		\centering
		\caption{Arguments for the higher activity of fresh or spent fuel. \linebreak \footnotesize{A checkmark indicates the presence of a code in an interview and an X indicates its absence.}}
		\label{table:Activtity of fresh and spent nuclear fuel}
		
		\begin{tabularx}{\linewidth}{ l XXXXXXXXXXXXX r}
			\toprule
			& A & B & C & D & E & F & G & H & I & J & K & L & M & $\Sigma$ \\
			\specialrule{0.5pt}{0.25pt}{0.25pt}
			\thead{Energy \\ (K-8) \hspace*{7em}}	& \checkmark & \checkmark & \checkmark & \checkmark & \checkmark & \checkmark & \checkmark & \checkmark & \checkmark & \checkmark & \checkmark & \checkmark & \checkmark & 13 \\
			\specialrule{0.3pt}{0.25pt}{0.25pt}
			\thead{Follow-up reactions \\ (K-9)} &   \checkmark & \checkmark & X & X & \checkmark & X & X &\checkmark  & \checkmark & \checkmark & X & X & \checkmark & 7 \\
			\specialrule{0.3pt}{0.25pt}{0.25pt}
			\thead{Absorption of energy \\ (K-10)} &  X  & X & X & \checkmark & \checkmark & X & X & X & X & X & \checkmark & \checkmark & X & 4 \\
			\specialrule{0.3pt}{0.25pt}{0.25pt}
			\thead{Multiplication by fission \\ (K-11)} &  \checkmark  & X & \checkmark & \checkmark & \checkmark & X &X & X & X & X & X & X & X & 4 \\
			\specialrule{0.3pt}{0.25pt}{0.25pt} 
		\end{tabularx}
	\end{table}
	
	\subsubsection{Interpretation}

	In describing the operation of nuclear power plants, it is clear that pre-service teachers need to differentiate more adequately between nuclear decay and nuclear fission. The boundaries in the description of by what means nuclear reactors obtain their energy are indistinct. However, this differentiation is imperative when describing byproducts, reactor safety, and associated safety measures. The importance of nuclear decay in providing energy in nuclear power plants is attributed to the decay of uranium (K-5):

	\begin{quote}
		A:\emph{ The uranium decays via some decay and their energy is released and then you have a radioactive daughter nucleus and that decays again via some decay where energy is released again and so on and so forth. That is just this decay chain. [...] altogether more energy is released than if only the uranium nucleus [...]. Because just by the many decays, if one adds that up, just the energy is bigger in the final effect.} (Anja, pos. 99)	
	\end{quote}
	
	\noindent Decay chains often accompany the reference to nuclear decay. However, the distinction between the chain reaction of neutron-induced nuclear fission and the decay chains is unclear. At this point, we want to point out that the ``decay chain'' and ``chain reaction'' differ in the German language (decay chains are called ``decay series''). While discussing the decay of uranium as a source of energy in nuclear power plants, pre-service teachers need to differentiate consistently between fission and decay. For example, this missing differentiation can be seen when the formation of two bodies is referenced (K-5 and K-7):

	\begin{quote}
		H:\emph{ [...] So. I don't know exactly what material is used for that, I mean uranium, but I'm not sure. When there's a part like that decays, it sort of breaks into these two new bodies and there's a bit of mass lost in the process. So the two end products weigh less than the sum, than the previous product. And this ``loss of mass'', in quotation marks, that is simply released as energy [...].} (Hannes, pos. 59)
	\end{quote}
	
	\noindent Besides the descriptions of nuclear decay or nuclear fission, the mass defect is also described here as playing a role in the operation of nuclear power stations. They identify the change in the observable mass as an energy-providing process, and the meaning of the equal sign in Einstein's formula is thus interpreted as a characteristic feature of nuclear processes. This is misleading. In nuclear physics, the mass defect is a significant phenomenon in which mass is ``lost''. The binding energies can be measured without a calorimeter by determining the nuclear masses in reactions or decays. However, a change in rest mass occurs during coal combustion or similar and is not a unique feature of nuclear reactions. Due to the scale of binding energies, however, it is observable here.
	
	The formation of two new bodies of approximately the same size indicates nuclear fission and not decay. The description of nuclear fission as a synonymous term for nuclear decay and the association of energy with the increased activity of spent nuclear fuel show a strong connection with the conceptualization of energy as the ``fuel'' of radioactivity. In the description of the energy produced from decay (K-5) and the potential explanation for higher activity of spent fuel, it is recognizable that Anja and Christopher consider a multiplication of nuclei. However, this idea did not play a role in their considerations of the energy supplied by nuclear reactors (K-11). Overall, pre-service teachers do not adequately differentiate the concepts of fission and decay. Pre-service teachers thereby syncretically combine concepts. Based on the collected data, context-dependency can only be assumed since the underlying conceptions remain vague.

	Pre-service teachers link radioactivity, activity, or the strength of radioactivity to a central conceptualization of energy. At this point, one can speak in a highly simplified way that pre-service teachers consider radioactivity and energy to be synonyms. Pre-service teachers suggest that the increased activity of used fuel elements may be due to multiplication by fission and other unspecified byproducts. The radioactive characteristic of the nucleus is supposed to be passed down to the daughter nuclei formed, which can sometimes lead to side reactions. Let us compare the codes M-2 (Rejection of activation) and K-10 (Absorption of energy) (see Table VII). We can also see a few cases where an activation hypothesis is denied at the beginning of the interview (\emph{e.g.} in the irradiation of food) but reemerges when pre-service teachers think about a possible higher activity of spent nuclear fuel. This can be directly traced back to different models of ionizing radiation (see D and L in \ref{Models,Interpretation}).

	\begin{table}[htp]	
		\centering
		\caption{\raggedright{Comparison of M-2, M-3, and K-10. \linebreak \footnotesize{A A checkmark indicates the presence of a code in an interview and an X indicates its absence.}}}
		\label{table:Comparison}
		\begin{tabularx}{\linewidth}{ l XXXXXXXXXXXXX r}
			\toprule
			& A & B & C & D & E & F & G & H & I & J & K & L & M & $\Sigma$ \\
			\specialrule{0.5pt}{0.25pt}{0.25pt}
			
			\thead{Rejection of activation \\ (M-2) }&    X         & X & X & X & \checkmark & X & \checkmark &X  & X & \checkmark & \checkmark & \checkmark & \checkmark & 6 \\
			\specialrule{0.5pt}{0.25pt}{0.25pt}
			\thead{Activation \\ (M-3)} &  \checkmark  & \checkmark & \checkmark & \checkmark & X & \checkmark & X & X & \checkmark & \checkmark & X & \checkmark & X & 8 \\
			\specialrule{0.5pt}{0.25pt}{0.25pt}
			\thead{Absorption of energy \\ (K-10)} &  X  & X & X & \checkmark & \checkmark & X & X & X & X & X & \checkmark & \checkmark & X & 4 \\
			\specialrule{0.3pt}{0.25pt}{0.25pt}
		\end{tabularx}
	\end{table}

	In the context of Chi's theory of conceptual change, the processes of decay and fission can be seen as examples of emergent and direct processes, respectively. Decay is an emergent process that cannot be directly controlled. In contrast, neutron-induced fission can be understood as a direct-controlled process. Additionally, spontaneous fission can be understood as an emergent process. Spontaneous fission occurs when a nucleus spontaneously splits into two or more smaller nuclei without being induced by an external particle or radiation.

	However, the coordination class of energy can also supplement this view by providing a framework for understanding how energy is involved in these processes. For example, in the case of decay, energy is released due to the spontaneous breakdown of a nucleus. This release of energy can be seen as an emergent process that cannot be directly controlled. However, the coordination class of energy can help us understand this role and how it affects the stability of the nucleus. Similarly, in the case of neutron-induced fission, energy is released when a neutron collides with a nucleus, causing it to split into two or more smaller nuclei. This direct process can be controlled, as it requires an external particle to induce fission. Once again, the coordination class of energy can help us understand how the energy released in this process affects the stability of the nuclei involved.

	In conclusion, while Chi's theory of conceptual change provides a valuable framework for understanding the processes of decay and fission, the coordination class of energy can supplement this view by providing a deeper understanding of the role of energy in these processes. By considering the role of energy in these processes, we can gain a more complete understanding of how they occur and how they can be controlled. One example of how complex energy can be as a concept is looking at the average binding energy of nucleons. While fission could be exothermic from iron onwards, this fact alone is insufficient to explain fission. Only nuclides from Th-232 and onwards potentially undergo fission at all. Therefore, more than simply looking at the average binding energy of nucleons in a nuclide, it is required to determine the nuclides fissile properties. Other factors, like transition states linked to potential energy, must also be studied.

	The coordination class of energy can provide a helpful framework for pre-service teachers to understand the differences between fresh and spent nuclear fuel. Our study observed that pre-service teachers always attribute the possible higher activity of fresh nuclear fuel to its greater energy. To better understand this, pre-service teachers could be introduced to the concept of energy as a coordination class, which would help them understand how energy is involved in decay, fission, and neutron activation processes. In addition, this could help them understand how the storage of spent nuclear fuel can increase its activity, as the radioactive isotopes created by neutron activation continue to undergo radioactive decay over time. Pre-service teachers must create more precise and knowledgeable explanations of spent nuclear fuel to enhance their comprehension of the coordination class of energy and its involvement in nuclear processes. This is essential to enable them to communicate the advantages and disadvantages of nuclear energy effectively and make well-informed decisions regarding its contribution to the energy industry.

	\section{CONCLUSION}\label{CONC}
	
	We were able to show that ``student conceptions'' regarding radioactivity are not limited to students but also extend to pre-service teachers, corroborating what studies have shown for other topics. Additionally, we are able to highlight the impact of context and the role of energy in understanding these concepts, which potentially bears implications for science education.

	\subsection{Conceptions of pre-service teachers about radioactivity}
	
	Our first research questions was concerned with which student conceptions could be observed in pre-service teachers. Following previously documented high-school student conceptions, pre-service teachers differentiate in part inadequately between radioactive matter and ionizing radiation. Some also fail to distinguish sufficiently between the processes of nuclear fission and nuclear decay. As a result, pre-service teachers misuse central terms related to the processes that occur in neutron-induced nuclear fission. The concepts of energy and size guide their concept of the interaction of ionizing radiation with matter, resulting in a naive, geometric understanding of the effective cross-section. Their understanding of the effect of ionizing radiation on biological structures is based on the damage to hereditary information by the energy input, which is connected to the concept of energy. A small proportion describes the effect as a stochastic phenomenon. The destruction of organs or genetic information is a direct process or input of energy in the context of the pre-service teachers' conceptions. Our findings on how pre-service teachers perceive radioactive matter and ionizing radiation are consistent with established research on understanding radiation-related concepts among students and the general public.
	
	It has become evident that Colclough's findings \cite{Colclough.2011} that pre-service teachers hold various student-like conceptions remain significant.  The struggles of pre-service teachers in this regard have important implications for science education. Inadequate differentiation between radioactive matter and ionizing radiation can impact the quality of instruction and hinder students' ability to grasp these complex concepts.

	\subsection{Context dependencies}

	Concerning the different scenarios (RQ2) and their possible impact on pre-service teachers conceptions, we showed that context-dependency is essential when students consider food irradiation or must explain the ``radioactive exposure'' of wild boars in southern Germany. The irradiation of food is often associated with the activation of the food. At the same time, the description of ``radioactive exposure'' in wild boars mainly relies on describing the transport of radioactive matter. We also showed that even if pre-service teachers negate the activation, it can reemerge in specific contexts (\emph{e.g.}, the activity of spent fuel or irradiation of food). The context-dependency and the reemergence of the activation concept are present in the priorly discussed quotes from Lars, which align with the observations presented in Table II. Lars noted that students tend to believe, falsely, that radioactivity leads to an object becoming radioactive (Lars, pos. 35). He later states that a radioactive dose received by a human is detectable as radioactive radiation (Lars, pos. 48-49).
	These quotes from Lars exemplify the data in Table II, highlighting the context-dependency and the reemergence of the activation concept in pre-service teachers' understanding.

	Specific characteristics of nuclear fission, like the creation of two nuclei of similar sizes, are merged with distinct aspects of decay, such as the chance of a decay chain, in the portrayal given by some pre-service teachers (like Georg). The finding that pre-service teachers' understanding of radiation and nuclear concepts can be influenced by context, even if they initially negate certain aspects (such as the activation of food), further illustrates the importance of context and prior beliefs in shaping individuals' understanding of these topics. This highlights the need for educators to carefully consider the context in which they teach radiation and nuclear concepts and to be aware of the potential for prior conceptions to influence student learning.

	\subsection{\emph{Energy} as a coordination class}
	
	The energy related to radioactive substances depends on their nuclear composition, which can be altered through decay or fission. It is crucial to understand the role of energy as a coordination class in these processes because of how it relates to other quantities like activity. To fully comprehend energy's involvement, clarifying its role in various interactions is essential. The activation concept can be explained through the coordination class of energy since energy is linked to its conservation. This complex concept structure is necessary to understand ionizing radiation and its interactions adequately. Pre-service teachers consistently focused on energy in their discussions on radioactivity, as evidenced by their recognition of the potential decrease in activity in spent nuclear fuel due to its energy content and their consideration of the penetrating power of ionizing radiation. Energy is a critical reference point for pre-service teachers' comprehension of radioactivity.

	\subsection{Limitations}

	While we conducted our research with pre-service teachers, it would be necessary to conduct this research with high school students since the conceptions of pre-service teachers are likely to be present in school students as well. We suggest further exploration of the possible relationship between fission and decay in different contexts, particularly for school students. This can be done through various prompts concerning nuclear weapons, nuclear reactors, and radioisotope batteries. Spent nuclear fuel and its properties also represent an interesting access point to nuclear physics and the problems of our modern world. This study will contribute decisively to describing the ``conceptual ecology'' of radioactivity or nuclear physics. An extension to teachers, in general, is also recommended.

	Although the study provides valuable insights into pre-service teachers' conceptions of nuclear physics, it has some limitations. These include its focus on a specific geographical and cultural context, a small sample size, and a lack of investigation into teaching methods and the impact of student characteristics. To overcome these limitations, future research could involve a more diverse population, larger sample sizes, and quantitative measures to supplement our qualitative findings. Furthermore, exploring teaching strategies and the influence of student characteristics could result in more inclusive and effective education in nuclear physics. The poor understanding of the effective cross-section is a good example of this importance for nuclear physics courses. Although cross-sections were the subject of the nuclear physics courses attended by the pre-service teachers interviewed, pre-service teachers still have only a simplified, geometric image. A comparison with computer games is a helpful analogy for understanding this context. These often have a hitbox for the execution of any combat action. The hitbox can be compared to an effective cross-section, which is currently programmed in non-intuitive ways in nuclear physics. We also know that such a simple analogy does not yet solve deeply rooted problems. Nevertheless, a calculation of the cross-section of action differs from understanding it. However, the goal of all coursework must be to describe the use of an energy concept and its limitations.

	\subsection{Future directions}
	
	Moving forward, teacher education programs must consider these findings, integrating comprehensive training on radioactivity and ionizing radiation concepts. By doing so, we can better prepare pre-service teachers to facilitate a deeper understanding of these topics among their students. Additionally, future research in this area should consider the influence of prior beliefs and values, as highlighted by Cooper's study \cite{Cooper.2003}, and explore effective instructional strategies to address these challenges. The body of literature discussed in this paper highlights the urgency of addressing the knowledge gaps and misconceptions surrounding radioactivity and ionizing radiation among pre-service teachers, ultimately enhancing the quality of science education and fostering a deeper understanding of these crucial scientific concepts.

	Levrini's study on how students learn from multiple contexts and definitions focuses on the ``proper time'' coordination class in teaching special relativity \cite{Levrini.2008}. Levrini's research method involved a mixed-methods approach to investigate how students learn the concept of proper time in different contexts and definitions. To advance the study on pre-service teachers' understanding of radioactivity, future research could explore how students learn from multiple contexts and definitions and how these experiences shape their understanding of the coordination class of energy concerning radioactive substances. This could involve examining how different teaching methods and contexts impact students' understanding of the role of energy in radioactivity-related processes, such as nuclear decay and fission, and how these concepts relate to other quantities like activity. By taking a similar approach to Levrini's study, future research could provide insights into how students learn about complex scientific concepts across multiple contexts and how this learning can be optimized through effective teaching practices.

	In the context of teacher training, developing a teaching-learning laboratory can help prospective teachers mirror their own conceptions and experience a confrontation with them while working with high school students. Moreover, since we found student conceptions in pre-service teachers, the interaction in the teaching-learning lab contributes to a better understanding of high school students' behalf and a better understanding of pre-service teachers' own knowledge constructions.

	\section*{AUTHOR DECLARATIONS}
	The authors have no conflicts to disclose.

	\begin{acknowledgments}
		This research was also funded by the \emph{Bundesministerium für Bildung und Forschung} (Federal Ministry of Education and Research), at the Professional School of Education Stuttgart Ludwigsburg in the project
		\emph{Lehrerbildung PLUS}, grant number: 01JA1907A. This project is part of the \emph{Qualitätsoffensive Lehrerbildung}, a joint initiative of the Federal Government and the states which aims to improve the quality of teacher training. The authors also like to thank Erich Starauschek for his helpful suggestions and discussions for this project.
	\end{acknowledgments}
	
\end{document}



\section*{Appendix A: Interview guide}

The following interview guide provides an overview of the topics addressed. When conducting a semi-structured interview, the order roughly follows this scheme.

\begin{itemize}
	\item Introduction
	\begin{itemize}
		\item Clarification of voluntariness, reference to data privacy.
		\item General introduction of the interviewer.
		\item ``Therefore, I would like to know what students think about radioactivity, so I ask for your help.''
		\item Depending on the choice of words of the interview partner, the same terms are used (e.g. radioactive radiation instead of ionizing radiation or similar)
	\end{itemize}
\end{itemize}

\begin{itemize}
	\item Description of radioactivity
	\begin{itemize}
		\item How would you explain to a fellow student who is not studying physics what the word radioactivity means?
		\begin{itemize}
			\item How do you imagine radioactivity yourself?
			\item How do you imagine the fission or decay?
		\end{itemize}
		\item Do you know any radioactive materials?
		\begin{itemize}
			\item Do you know of some radioactive materials that we might encounter in our everyday lives?
			\item What distinguishes for you a radioactive substance from one that is not radioactive?
		\end{itemize}
		\item How would you describe the ``strength'' of the radioactivity?
		\item How would you describe the ``abundance'' of radioactivity? Assuming we had 1 kg of this or that substance
		\item What does the half-life for a single atomic nucleus mean to you?
	\end{itemize}
\end{itemize}

\begin{itemize}
	\item Ionizing radiation
	\begin{itemize}
		\item Where do we encounter radioactive radiation?
		\item Do you know different types of radiation?
		\begin{itemize}
			\item What do you imagine alpha, beta or gamma radiation to be?
			\item In contrast, how do you imagine (non-radioactive)* radiation?
		\end{itemize}
		\item How would you describe the effects of radioactive radiation on humans?
		\begin{itemize}
			\item What effects does this ionization have on, for example, humans?
			\item How do you imagine the change of the genome/DNA or organs?
		\end{itemize}
		\item How would you describe the effects of radioactive radiation on non-living things?
	\end{itemize}
\end{itemize}

\begin{itemize}
	\item ``In a typical illustration, the penetrating ability of different types of (ionizing/radioactive) radiation is shown.'' (Prompt 1)
	\begin{itemize}
		\item How do you explain the different behavior of the different types of radiation?
		\begin{itemize}
			\item What physical properties can you deduce from this experiment?
			\item Is it also possible to derive properties from this experiment as to what kind of threat the individual species pose to us?
		\end{itemize}
	\end{itemize}
\end{itemize}

\begin{itemize}
	\item ``In the food industry, food is irradiated to improve shelf life and for hygienic reasons.'' (Prompt 2)
	\begin{itemize}
	\item How do you imagine this irradiation?
	\item How would you imagine the benefit of such irradiation?
	\item Do you see any technical advantages to using the procedure this way?
	\item You talked about \emph{residues/changes} of the chili. Could you elaborate on that again?	
	
\newpage
	
\end{itemize}
\item ``One also uses radioactive drugs, so-called radiopharmaceuticals, in medicine. Or also a so-called radiation therapy.'' (Prompt 3)
\begin{itemize}
\item How do you imagine the benefits or effects of radiopharmaceuticals?
\item What disadvantages do you see in the use of radiopharmaceuticals?
\end{itemize}
\item Nuclear power plants
\begin{itemize}
\item How do you imagine a nuclear power plant works?
\begin{itemize}
	\item Could you go into more detail about how it is used?
	\item How do you imagine nuclear fission?
\end{itemize}
\item What advantages do you see in using nuclear energy?
\item What disadvantages do you see in using nuclear energy?
\item Could you go into more detail about the term repository/hazard?		
	\end{itemize}
\end{itemize}

\begin{itemize}
	\item „It has been 33 years since the disaster in Chernobyl, but the effects of the reactor accident are still measurable in this country. In some regions of southern Germany, every fifth wild boar is contaminated with radioactivity. Their meat may not be sold.“ (Prompt 4)
	\begin{itemize}
		\item How would you describe the effects of the accident in Chernobyl?
		\item Based on the Chernobyl accident, how would you describe the history from the accident to the radioactive contamination of wild boars in southern Germany?
		\item How would you describe the problems of radioactive waste storage?
		\begin{itemize}
			\item What do you think suggests that the fuel could be more radioactive?
			\item And what is your argument that the waste could be more radioactive?
		\end{itemize}
	\end{itemize}
\end{itemize}

\begin{itemize}
\item {Sounding off}
\begin{itemize}
\item ``We've been talking about radioactivity for a long time now. Do you want to add anything else?''
\item ``Do you have any questions about radioactivity?''
\item ``Thank you very much for the conversation!''
\item General questions (age, gender, degree, number of semesters)
	\end{itemize}
\end{itemize}



\section*{Appendix B: Codebook}

\noindent In addition to the abbreviated designations of the codes (e.g. K-5), the names of the codes and their explanations are recorded in the coding book. The anchor examples are taken from the interviews. The interviews were held in German and are available upon proper request. The short names of the codes are used in the second part of the appendix (Appendix C).

\begin{table}[h!]	
	\centering
	\caption{Codebook (Part a)}
	\footnotesize{
		\begin{tabular}{  m{0.06\linewidth}  m{0.20\linewidth} m{0.275\linewidth}  m{0.375\linewidth}  } 
			\hline
			\textbf{Code} & \textbf{Name} & \textbf{Explanation} & \textbf{Anchor example} \\
			\midrule
			R-1* & Instability &  Requirement for the occurrence of radioactivity lies in the instability of the atom or atomic nucleus.& \glqq I would say that radioactivity means that there are unstable atomic nuclei which decay. And that there are different decays. So, that it is a process that takes place in the atomic nucleus.\grqq{} (Elena, pos. 3)\\
			\midrule
			R-2* & Emission of particles& Radioactivity occurs when particles are emitted during a nuclear decay. & \glqq So I would say that radioactivity occurs when an atomic nucleus decays. Exactly, then particles are emitted and then one speaks of radioactivity.\grqq{} (Anja, pos. 3)\\	
			\midrule
			R-3* & Emission of ionizing radiation/energy & Radioactivity involves the release of ionizing radiation or energy. & \glqq Radioactivity, I would have explained now like this, is an expression for the fact that something radiates by atomic decay processes. [...] unstable atomic nuclei or unstable atoms emit this radioactive or this ionizing radiation.\grqq{} (Georg, pos. 3) \\
			\midrule
			R-4* & Transmutation & Radioactivity leads to the appearance of altered nuclei or other elements. & \glqq Exactly, so these are just reactions. The nucleus, i.e. this one element, becomes another.\grqq{} (Ina, pos. 5)\\
			\midrule
			R-5* & Ionization & Radioactivity is characterized by the ionization of the receiving material.	& \glqq Ionizing radiation is simply radiation that manages to knock electrons out of atom or out of molecules.\grqq{} \linebreak (Dalia, pos. 3)\\
			\midrule
			R-6* & Effects on non-living matter.	& The effect of (ionizing) radiation on non-living matter is considered possible in principle.	& \glqq I can imagine that, I mean, when helium nuclei fly through the air and then hit another object, that they are absorbed somehow or. Or that they change something in the material. When they hit there. For sure yes. [...] \grqq{} (Konstantin, pos. 31)\\
			\midrule
			R-7* & Radioactive elements (Z>84)	& Examples of radioactive materials are elements with an atomic number greater than 84.	& \glqq I don't know, I don't know that well either. I think it's just these, you always see these plutonium or uranium rods there.\grqq{} (Felix, pos. 79)\\
			
			\midrule
			R-8* & Radioactive \glqq elements\grqq{} (Z<85)	& Isotopes of certain elements. Often with the link to occurrences of radioactive isotopes of some elements and their condition for radioactivity.	& \glqq So an element can be radioactive, can be radioactive in certain composition, can't it?\grqq{} (Christopher, pos. 21) \\
			\midrule
			R-9* & Natural occurrence	& Ionizing radiation also occurs in natural processes. Name everyday objects that release ionizing radiation.	& \glqq Basically, [radioactive radiation] is always there. It is just usually a very low concentration.\grqq{} (Georg, pos. 25) \\
			\midrule
			R-10*& Three types of ionizing radiation & It is explicitly formulated that there are three different types of ionizing radiation.& \glqq Exactly, then I would, so yes, there are three different types of decays.\grqq{} (Anja, pos. 5) \\
			\midrule
			R-11* &	Alpha radiation & Description of the alpha radiation (excl. description of the penetrability) & \glqq So now with alpha radiation this would be just such a helium nucleus, thus speak two protons and two neutrons.\grqq{} (Dalia, pos. 5) \\
			\midrule
			R-12* & Beta radiation & Description of beta radiation (excl. description of penetrability) & \glqq Beta radiation are electrons, if I still have it right in my head. Exactly.\grqq{} (Konstantin, pos. 3) \\
			\midrule
			R-13* & Gamma radiation & Description of gamma radiation, including its \glqq dissimilarity\grqq (excl. the description of the penetrability) & \glqq  [A]nd with the gamma decay it is practically no particle, but just such an energy. Then one says then also this gamma or so, gamma particle one says then to it.\grqq{} \glqq Yes, gamma radiation are effectively quasi electromagnetic waves, so photons, which are emitted when, when I release quasi energy in a nuclear decay process.\grqq{} \linebreak (Anja, pos. 5) \\
			
			\bottomrule
			
		\end{tabular}
	}
\end{table}

\FloatBarrier

\begin{table}[h!]	
	\centering
	\caption{Codebook (Part b)}
	\footnotesize{
		\begin{tabular}{  m{0.06\linewidth}  m{0.20\linewidth} m{0.275\linewidth}  m{0.375\linewidth}  }  
			\toprule
			\textbf{Code} & \textbf{Name} & \textbf{Explanation} & \textbf{Anchor example} \\
			\midrule
			R-14* &	Strength of radioactivity is described by energy content& The strength of radioactivity is described by the energy content of the radiation.	& \glqq Yes, so to speak, the wavelength of the, of the radiation that. Yes, so the energy content of the radiation, which is then the measure of how strong somehow the radiation is.\grqq{} \linebreak (Benjamin, pos. 15) \\
			\midrule
			R-15* & The strength of the radioactivity is described by the activity & The strength of the radioactivity is described by the number of decaying nuclei per time unit. & \glqq So there are relatively many different quantities for it. So one example would be the activity. This is a quantity, which is given in Becquerel, and it indicates, for example, how many decays take place in a certain time interval.\grqq{} (Dalia, pos. 11)  \\
			\midrule
			R-16* &Macroscopic interpretation of the half-life & The half-life  time describes the decay of half of a measurable quantity.	& \glqq So half-life describes the time it takes for half of the nuclei present at the beginning to decay.\grqq{} (Anja, pos. 23)\\
			\midrule
			R-17* & Microscopic-statistical interpretation of the half-life & The half-life describes the expected value for the decay of a single atomic nucleus.	& \glqq So it is now, one cannot say now in three seconds it decays already. It's a bit random, isn't it?\grqq{} \linebreak (Christopher, pos. 19)\\
			\midrule
			D-1 & Penetration capability allows risk to be assessed & The penetration capability of different types of radiation is directly linked to the hazard of the respective type of radiation. Supplemental category for uptake and associated hazard. & \glqq So [...] Exactly. One sees practically that if we are hit or confronted with alpha or beta radiation from the outside, that it is possible to protect oneself with lead for example or with aluminum. And just with gamma radiation, one cannot protect oneself with lead.\grqq{} (Anja, pos. 51) \\
			\midrule
			D-2 & Risk assessment in case of uptake into the body & In case of uptake into the body, however, the description by means of the penetration ability is no longer given. & \glqq But it is also so that if one takes up now for example alpha radiation or any alpha - thus any radioactive particles, which decay alpha, into the body, that it is then just much more dangerous than for example gamma radiation, because, because the just much, because the range is then just nevertheless large enough and all organs are hit.\grqq{} \linebreak (Anja, pos. 51) \\
			\midrule
			D-3 & Size & The size (volume or similar) decides on its penetrability. Excluded from this are statements which refer exclusively to a simplification. & \glqq And then the second one, the light green one, is somehow beta, beta radiation. Because just the beta particle, thus the proton or electron is just smaller and therefore not so strongly by impact processes somehow energy loses. But still it is shielded by aluminum.\grqq{} (Anja, pos. 45) \\
			\midrule						
			D-4 & Energy & The energy content of the respective type of radiation determines its penetration capability. & \glqq That they have different energies. [...] Well, alpha radiation less energetic, beta radiation in the middle and gamma radiation just the highest energetic.\grqq{} (Dalia pos. 39-41) \\
			\midrule
			D-5 & Interaction & Interaction of the particles or similar decides about their penetrability (excluding purely mechanical interactions). & \glqq Because that strongly interacts with other particles, thereby just loses energy and thereby just also then doesn't move on.\grqq{} (Anja, pos. 45)  \\
			\midrule
			D-6 & Density of absorber & The density of the absorber decides on the penetration ability with & And lead is known for its high density and also for its heavy nuclear charge number. This top beam, this radioactive, this gamma beam, is then already strongly attenuated by this.\grqq{} \linebreak (Konstantin, pos. 33) \\
			\bottomrule
		\end{tabular}
	}
\end{table}

\FloatBarrier
\newpage

\begin{table}[h!]	
	\centering
	\caption{Codebook (Part c)}
	\footnotesize{
		\begin{tabular}{  m{0.06\linewidth}  m{0.20\linewidth} m{0.275\linewidth}  m{0.375\linewidth}  }  
			\toprule
			\textbf{Code} & \textbf{Name} & \textbf{Explanation} & \textbf{Anchor example} \\
			\midrule
			M-1 & Transport of  material & Radioactivity is distributed by the transport of radioactive material. & \glqq There has burned down a nuclear power station at that time and the, the materials have been transported over the atmosphere up to us. But what exactly went wrong at that time, no. I'm sure I read that at some point, but I didn't remember that ((laughs)) in detail.\grqq{} (Benjamin, pos. 77) \\
			\midrule
			M-2 & Exclusion of activation & Explicit exclusion of the storage of \glqq radiation \grqq{} & \glqq My if I irradiate a chili, then it does not become a radiator itself. But it is then not dangerous for someone holding the package, because it itself does not radiate. And that is not so that the chili could store the radioactivity.\grqq{} (Konstantin, pos. 45) \\
			\midrule
			M-3 & Activation & Activation by absorption into the absorber & \glqq Yes, that something changes in the chilies. And that something changes in the sense of, well, they are later eaten, and that then some strange changed atoms are absorbed by the body. [...] radiation.\grqq{} (Dalia, pos. 65) \\
			\midrule		
			M-4 & Transport of radioactivity & Transport of radioactivity or radiation & \grqq So I am just thinking. So generally it was always said that about the precipitation, the radiation gets then into the atmosphere and comes then just in the thunderclouds and is then rained down over the rest of Europe. Depending upon how now also the winds stand. I ask myself also however again, which of it becomes there now quasi physically.\grqq{} (Jan, pos. 77) \\
			\midrule
			B-1 & Biological structures &Damage to biological structures (excl. damage to biochemical structures such as DNA or similar) & \glqq So in the case of radiation therapy, we know this from cancer in particular. You get chemotherapy and sometimes also radiation therapy. And the benefit, I think you just try to kill the cancer cells targeted. And exactly tries to ensure that the cancer does not spread so far.\grqq{} (Konstantin, pos. 51) \\
			\midrule
			B-2 & Biochemical structures & Naming specific changes to DNA, genetic material or similar. & \glqq I think the DNA is damaged.\grqq{} \linebreak (Elena, pos. 61) \\
			\midrule
			B-3 & Stochastic effect & The biological effect of ionizing radiation is not a deterministic process (respectively the development of cancer) & \glqq I have locally an effect, but it is nevertheless also somehow a stochastic process that I do not know at all so exactly, what my radiation with the, yes cells then yes in the final effect then more exactly will cause. What exactly will come out in the end. At least not one hundred percent. I think a large part will actually be burned. But in the peripheral area probably not.\grqq{} \linebreak (Benjamin, pos. 55) \\			
			\bottomrule
		\end{tabular}
	}
\end{table}

\FloatBarrier

\newpage

\begin{table}[h!]	
	\centering
	\caption{Codebook (Part d)}
	\footnotesize{
		\begin{tabular}{  m{0.06\linewidth}  m{0.20\linewidth} m{0.275\linewidth}  m{0.375\linewidth}  }  
			\toprule
			\textbf{Code} & \textbf{Name} & \textbf{Explanation} & \textbf{Anchor example} \\
			\midrule
			K-1* & Socioecological aspects & Reference so socioecological aspects of the application of nuclear energy & \glqq But in the sense of sustainability of course now thought in the short term, is, of course, climate change is actually the climate catastrophe, one should better say, an urgent problem. And yes, nuclear power plants, construction phase and all, already have lower CO2 emissions than coal-fired power plants. That cannot be denied.\grqq{} \linebreak (Benjamin, pos. 65) \\
			\midrule
			K-2* & Control of nuclear power plants & Description of possible control mechanisms in nuclear power plants & \glqq Well, you also have to make sure that it is cooled, because the radioactive decay also releases heat. And that, it is somehow so, if it, the warmer it becomes, the more dangerous it is. Because, then it can also come to this core meltdown. And then, if it were left as it is, I think it would simply develop such a high heat that it would melt through the concrete wall. And that's why it has to be cooled somehow.\grqq{} \linebreak (Anja, pos. 95) \\
			\midrule
			K-3* & Final disposal of nuclear waste & Description of final disposal and its requirements & \glqq Yes, that is always this big problem also in society, when they say, what do we do with our waste. What remains in the end is the whole outside of the nuclear reactor. It has to be stored somewhere and it will continue to radiate for a while.\grqq{} (Felix, pos. 89) \\
			\midrule
			K-4* & Possibility of accidents & descriptions of accidents at nuclear power plants and their occurrence & \glqq Well. A very, very practical problem: It is always stupid when a nuclear power plant explodes.\grqq{}  \linebreak (Benjamin, pos. 63)\\
			\midrule
			K-5 & Decay/radioactivity & Nuclear power plants use the decay of uranium to provide energy. & \glqq So, through the decay of uranium, heat is practically also released and the heat is then used to heat water [...] .\grqq{} (Anja, pos. 75) \\
			\midrule
			K-6 &  Fission & Nuclear energy is provided by the induced nuclear fission of uranium. Partwise decay is also mentioned here, but an initiating part is mentioned with. & \glqq So that's, as far as I remember, that it's a chain reaction, so each split nucleus gives off two more splitting particles. [...] Anyway, every particle that is split, it sort of causes another two to be split, and that then exponentiates.\grqq{} (Georg, pos. 75) \\
			\midrule
			K-7 & Mass defect & The mass defect is responsible for the release of energy.  & \glqq When such a part decomposes, it breaks into these two new partial bodies, so to speak, and a bit of mass is lost in the process. So the two end products weigh less than the sum, than the previous product. And this loss of mass, in quotation marks, that is simply released as energy and that is then what is called [...] .\grqq{} (Hannes, pos. 59) \\
			\midrule
			K-8 & Energy & The fuel is before more radioactive, because it can still provide energy, a critical mass or radiation. & \glqq Well, that just with such a, yes it radiates, and there then, so radiates then maybe less energetically or as far as the intensity is concerned.\grqq{} \linebreak (Christopher, pos. 128) \\
			\midrule
			K-9 & Follow-up reactions & The waste could be more radioactive due to new products. & \grqq That depends on the material again. Or? How active such a medium is. If the decay bodies then or these new elements. I don't know what there would be exactly related to the activity and the size or something for example.\grqq{} (Hannes, pos. 69) \\
			\midrule
			K-10 & Absorption of energy & The waste becomes more radioactive by a storage of energy or excitation. & \glqq I'd have to think about that for a moment [...] Maybe if you have nuclear fissions there. That goes perhaps into the right direction. If you somehow divert energy with this uranium by nuclear fission, that then later the material is more radioactive than before.\grqq{} (Konstantin, pos. 75) \\
			\midrule
			K-11 & Multiplication by fission & The waste could be more radioactive, because the number of still radioactive nuclei is multiplied by fission. & \glqq I would have said now that afterwards. Because nevertheless several atoms then have, because I split the nevertheless.\grqq{} (Dalia, pos. 85) \\
			\bottomrule
		\end{tabular}
	}
\end{table}

\FloatBarrier


\section*{Appendix C: Coding table}
	
	\noindent In the following tables, the presence of a code in an interview is marked with differently marked fields.  A checkmark indicates the presence of a code in an interview and an X indicates its absence. The last column shows in how many interviews a code could be found overall.

	\begin{table}[h!]	
		\centering
		\caption{Coding table}
		\normalsize{
			\begin{tabularx}{0.95\linewidth}{ l XXXXXXXXXXXXX r}
				\toprule
				& A  & B  & C  & D  & E  & F  & G  & H  & I  & J  & K  & L  & M  & $\Sigma$ \\
				R-1*   & \checkmark & \checkmark   &  \checkmark   & \checkmark   & \checkmark & \checkmark & X & \checkmark & \checkmark & \checkmark   & X & \checkmark & \checkmark & 11       \\
				R-2*   & \checkmark & \checkmark   &  \checkmark   & \checkmark   & X & X & \checkmark & \checkmark & X & \checkmark   & \checkmark & X & \checkmark & 9       \\ 
				R-3*   & X & \checkmark   &  \checkmark   & \checkmark   & \checkmark & \checkmark & \checkmark & X & \checkmark & \checkmark   & X & \checkmark & \checkmark & 10       \\
				R-4*   & \checkmark & X   &  \checkmark   & \checkmark   & X & X & X & X & \checkmark & \checkmark   & X & \checkmark & \checkmark & 7       \\
				R-5*   & \checkmark & X   &  X   & \checkmark   & \checkmark & \checkmark & \checkmark & X & X & \checkmark   & X & X & X & 6       \\
				R-6*   & \checkmark & \checkmark   &  X   & \checkmark   & X & \checkmark & \checkmark & \checkmark & \checkmark & \checkmark   & \checkmark & \checkmark & X & 10       \\
				\midrule
				R-7*  	& \checkmark & \checkmark  & \checkmark  & \checkmark  & \checkmark  & \checkmark  & \checkmark  & \checkmark  & X  & \checkmark  & \checkmark  & \checkmark  & \checkmark & 12 \\
				R-8*   & X & X   &  \checkmark   & X   & \checkmark & \checkmark & \checkmark & X & \checkmark & \checkmark   & X & X & X & 6       \\
				R-9*   	& \checkmark & X  & \checkmark  & \checkmark  & \checkmark  & \checkmark  & \checkmark  & \checkmark  & \checkmark  & X  & \checkmark  & \checkmark  & \checkmark & 11 \\
				\midrule
				R-10*   & \checkmark & X   &  \checkmark   & \checkmark   & \checkmark & X & \checkmark & X & X & X   & \checkmark & X & \checkmark & 7       \\
				R-11*   & \checkmark & X   &  \checkmark   & \checkmark   & \checkmark & \checkmark & X & \checkmark & X & X   & \checkmark & \checkmark & \checkmark & 9       \\
				R-12*   & \checkmark & \checkmark   &  \checkmark   & \checkmark   & X & \checkmark & X & \checkmark & \checkmark & \checkmark   & \checkmark & \checkmark & \checkmark & 11       \\
				R-13*   & \checkmark & \checkmark   &  \checkmark   & \checkmark   & X & \checkmark & X & \checkmark & \checkmark & X   & \checkmark & \checkmark & \checkmark & 10       \\
				\midrule
				R-14*   & X & \checkmark   &  X   & \checkmark   & X & \checkmark & X & X & X & X   & \checkmark & X & X & 4       \\
				R-15*   & \checkmark & \checkmark   &  X   & \checkmark   & \checkmark & \checkmark & \checkmark & \checkmark & \checkmark & \checkmark   & X & \checkmark & X & 10       \\
				R-16*   & \checkmark & \checkmark  & \checkmark  & \checkmark  & \checkmark  & \checkmark  & \checkmark  & \checkmark  & X  & \checkmark  & \checkmark  & \checkmark  & \checkmark & 12 \\
				R-17*   & \checkmark & \checkmark  & \checkmark  & \checkmark  & \checkmark  & X  & \checkmark  & \checkmark  & \checkmark  & \checkmark  & \checkmark  & \checkmark  & X & 11 \\
				\midrule
				D-1   & \checkmark & \checkmark  & \checkmark  & \checkmark  & \checkmark  & \checkmark  & \checkmark  & \checkmark  & \checkmark  & \checkmark  & \checkmark  & \checkmark  & \checkmark & 13 \\
				D-2   & \checkmark & \checkmark  & \checkmark  & \checkmark  & \checkmark  & X  & \checkmark  & X  & X  & \checkmark  & X  & X  & \checkmark & 8 \\
				\midrule
				D-3   & \checkmark & X   &  X   & X   & \checkmark & \checkmark & \checkmark & \checkmark & \checkmark & X   & \checkmark & \checkmark & \checkmark & 9       \\
				D-4    & X   & X   & \checkmark & \checkmark & X   &  X  & X   & \checkmark & \checkmark & \checkmark & \checkmark & X   & \checkmark & 7       \\
				D-5 & \checkmark & \checkmark & X   & X   & \checkmark & X   & \checkmark & \checkmark & \checkmark & X   & \checkmark & \checkmark & X   & 8       \\
				D-6 &  X  &  X  &  X  &  X  &  X  & \checkmark & \checkmark &  X  & X   & \checkmark & \checkmark & X   & X   & 4  \\	
				M-1	& \checkmark & \checkmark  & \checkmark  & X  & \checkmark  & \checkmark  & \checkmark  & \checkmark  & \checkmark  & \checkmark  & X  & X  & \checkmark & 10 \\
				M-2 &       X                  & X  & X  &  X & \checkmark  & X  & \checkmark  &X   & X  & \checkmark  & \checkmark  & \checkmark  & \checkmark & 6  \\
				M-3 &   \checkmark   & \checkmark  &  \checkmark & \checkmark  & X  &  \checkmark & X  & X  & \checkmark  & \checkmark  & X  & \checkmark  & X & 8  \\
				M-4 & X & X  & X  & \checkmark  & X  & X  & X  & X & \checkmark  & \checkmark  & X  & X  & X & 3  \\
				\midrule
				B-1	& \checkmark & \checkmark  & \checkmark  & \checkmark  & \checkmark  & \checkmark  & \checkmark  & \checkmark  & \checkmark  & \checkmark  & \checkmark  & \checkmark  & \checkmark & 13 \\
				B-2 &      \checkmark                  & \checkmark  & X  &  X & \checkmark  & \checkmark  & \checkmark  &\checkmark   & \checkmark  & X  & \checkmark  & \checkmark  & \checkmark & 10  \\
				B-3 &   X   & \checkmark  &  X & X  & X  &  X & \checkmark  & \checkmark  & X  & X  & X  & \checkmark  & X & 4  \\
				\midrule
				K-1*   & \checkmark & \checkmark   &  X   & X   & X & \checkmark & X & \checkmark & \checkmark & \checkmark   & \checkmark & \checkmark & \checkmark & 9       \\
				K-2*   & \checkmark & \checkmark   &  X   & \checkmark   & \checkmark & \checkmark & \checkmark & \checkmark & \checkmark & X   & X & \checkmark & \checkmark & 10       \\
				K-3*   & \checkmark & \checkmark   &  X   & \checkmark   & \checkmark & \checkmark & \checkmark & \checkmark & \checkmark & \checkmark   & \checkmark & \checkmark & \checkmark & 12       \\
				K-4*   & X & \checkmark   &  \checkmark   & \checkmark   & \checkmark & \checkmark & \checkmark & \checkmark & \checkmark & \checkmark   & \checkmark & \checkmark & \checkmark & 12       \\
				\midrule
				K-5	& \checkmark & X  & \checkmark  & X  & X  & \checkmark  & \checkmark  & \checkmark  & \checkmark  & \checkmark  & \checkmark  & X  & X & 8 \\
				K-6 &   X   & \checkmark  &  X & \checkmark  & \checkmark  &  \checkmark & \checkmark  & X  & X  & X  & X  & \checkmark  & \checkmark & 7  \\
				K-7 &      X                  & X  & X  &  X & X  & X  & \checkmark  &\checkmark   & X  & X  & X  & X  & X & 2  \\	
				K-8	& \checkmark & \checkmark  & \checkmark  & \checkmark  & \checkmark  & \checkmark  & \checkmark  & \checkmark  & \checkmark  & \checkmark  & \checkmark  & \checkmark  & \checkmark & 13 \\
				K-9 &      \checkmark  & \checkmark  & X  &  X & \checkmark  & X  & X  &\checkmark   & \checkmark  & \checkmark  & X  & X  & \checkmark & 7  \\
				K-10 &   X   & X  &  X & \checkmark  & \checkmark  &  X & X  & X  & X  & X  & \checkmark  & \checkmark  & X & 4  \\
				K-11 &   \checkmark   & X  &  \checkmark & \checkmark  & \checkmark  &  X &X  & X  & X  & X  & X  & X  & X & 4  \\
				\bottomrule		
		\end{tabularx}}
	\end{table}